\documentclass[superscriptaddress,nofootinbib,twocolumn,10pt]{revtex4-2}

\usepackage{amsthm,amsmath,amssymb}
\usepackage{mathrsfs}
\usepackage{appendix}
\usepackage{amstext}
\usepackage{graphicx}
\usepackage{url}
\usepackage{bm}
\usepackage{braket}
\usepackage{lineno}
\usepackage[usenames,dvipsnames]{color}
\usepackage[colorlinks=true,citecolor=blue,urlcolor=black]{hyperref}% add hypertext capabilities
\usepackage{dcolumn}% Align table columns on decimal point
\usepackage{natbib}%to cite
\usepackage{booktabs}
\usepackage{color}

\begin{document}
\title{Experimental Efficient Source-Independent Quantum Secret Sharing against Coherent Attacks}

\author{Yi-Ran Xiao}
\thanks{These authors contributed equally.}
\affiliation{National Laboratory of Solid State Microstructures and School of Physics, Collaborative Innovation Center of Advanced Microstructures, Nanjing University, Nanjing 210093, China}
\affiliation{School of Physics and Key Laboratory of Quantum State Construction and Manipulation (Ministry of Education), Renmin University of China, Beijing 100872, China}

\author{Hua-Lei Yin}
\email{hlyin@ruc.edu.cn}
\thanks{These authors contributed equally.}
\affiliation{School of Physics and Key Laboratory of Quantum State Construction and Manipulation (Ministry of Education), Renmin University of China, Beijing 100872, China}
\affiliation{National Laboratory of Solid State Microstructures and School of Physics, Collaborative Innovation Center of Advanced Microstructures, Nanjing University, Nanjing 210093, China}
\affiliation{Beijing Academy of Quantum Information Sciences, Beijing 100193, China}

\author{Wen-Ji Hua}
\thanks{These authors contributed equally.}
\affiliation{National Laboratory of Solid State Microstructures and School of Physics, Collaborative Innovation Center of Advanced Microstructures, Nanjing University, Nanjing 210093, China}
\affiliation{School of Physics and Key Laboratory of Quantum State Construction and Manipulation (Ministry of Education), Renmin University of China, Beijing 100872, China}

\author{Xiao-Yu Cao}
\affiliation{National Laboratory of Solid State Microstructures and School of Physics, Collaborative Innovation Center of Advanced Microstructures, Nanjing University, Nanjing 210093, China}
\affiliation{School of Physics and Key Laboratory of Quantum State Construction and Manipulation (Ministry of Education), Renmin University of China, Beijing 100872, China}

\author{Zeng-Bing Chen}\email{zbchen@nju.edu.cn}
\affiliation{National Laboratory of Solid State Microstructures and School of Physics, Collaborative Innovation Center of Advanced Microstructures, Nanjing University, Nanjing 210093, China}
\date{\today}

\begin{abstract}
Source-independent quantum secret sharing (SI QSS), while essential for secure multiuser cryptographic operations in quantum networks, faces significant implementation challenges stemming from the inherent complexity of generating and distributing multipartite entangled states. Recently, a resource-efficient SI QSS protocol utilizing entangled photon pairs combined with a postmatching method has been proposed to address this limitation.
In this Letter, we report an experimental demonstration of this protocol using high-fidelity polarization-entangled photon pairs in a star topology. For a three-user network, we obtain secure key rates of 21.18, 4.69, and 1.71 kbps under single-user channel losses of 7.6, 10.9, and 12.9 dB respectively. Furthermore, under conditions of equal channel loss per user, we achieve secure key rates of 6.97, 6.46, and 5.88 kbps for three-, four-, and five-user scenarios respectively. These results demonstrate the advantageous independence of the key rate from the number of users. Our work paves the way for large-scale deployment of SI QSS in multiuser quantum networks.
\end{abstract}

\maketitle

\section{Introduction}
Quantum communication~\cite{xu2020secure,yin2023experimental,azuma2023quantum,li2024asynchronous} has experienced significant advancement over the past few decades owing to its information-theoretic security advantages beyond classical communication.
Initially, quantum communication was primarily implemented in a point-to-point scenario, for example quantum key distribution~\cite{Zhou2023OvercomeRateloss,xie2022breaking,zeng2022mode}, ensuring secure encryption communication between two users.
With the rapid development of quantum information technology, building a large-scale quantum network~\cite{azuma2023quantum,kimble2008quantum,wehner2018quantum,li2024asynchronous} would extend the benefits of quantum communication to multiple users, going beyond point-to-point communication.

Secret sharing, first proposed independently by Shamir~\cite{Shamir1979SecretSharing} and Blakley~\cite{Blakley1979SecretSharing} in 1979, is a cryptographic primitive among multiple users in networks.
In secret sharing, a dealer splits a secret into several parts and distributes them among various players, and the secret can only be recovered when authorized subsets of players cooperate to share their parts of the secret.
As a combination of secret sharing and quantum mechanics, quantum secret sharing (QSS)~\cite{Hillery1999QSS,Cleve1999QSS,Fu2015MDIQSSQCKA,gu2021secure,wang2024experimental} offers a quantum-enhanced security against the eavesdroppers equipped with quantum computers and has become a notable component of quantum networks.

The first QSS protocol was introduced by Hillery \textit{et al.}~\cite{Hillery1999QSS} in 1999, which utilizes a three-photon Greenberger-Horne-Zeilinger (GHZ) state to share the secret.
Over the past two decades, significant theoretical ~\cite{lance2004tripartite,Singh2005GeneralizedQSS,Yu2008MultilevelQSS,Markham2008GraphQSS,Cai2017MultimodeGraphQSS,de2020experimental,Tavakoli2015DLevelQSS,Li2023AMDIQSS,zhang2024device} and experimental~\cite{schmid2005experimental,chen2005experimental,Gaertner2007FourPartyQSS,Bell2014ExperimentGraphQSS,lu2016secret,Zhou2018BoundEntangleQSS,Williams2019EntangleQSS,lee2020quantum,liu2021photonic,shen2023experimental,liu2023experimental,qin2024efficient} advances have been made, leading to the development and proof-of-principle demonstration of diverse QSS protocols.
Combined with other quantum cryptographic protocols, QSS plays a crucial role in enabling various quantum network applications, ranging from multiparty quantum computing~\cite{gottesman1999demonstrating} to quantum blockchain~\cite{Cao2024Qecommerce,jing2024experimental}.
Nevertheless, owing to the low generation rate and fidelity of multipartite entangled states under current experimental conditions~\cite{yao2012observation,wang2016experimental,zhong201812}, QSS protocols relying on such states face challenges in practical applications. 
Besides, if multiparticle correlation is simply utilized, there will still be internal participant attacks~\cite{karlsson1999quantum,qin2007cryptanalysis}.
Furthermore, numerous experimental studies~\cite{Zhou2018BoundEntangleQSS,Williams2019EntangleQSS,liu2023experimental,qin2024efficient} lack the composable security analysis under a finite-key regime against the coherent attacks.
Recently, a resource-efficient source-independent QSS (SI QSS) protocol~\cite{xiao2024source} has circumvented the generation and distribution of multipartite entangled states through the postmatching method~\cite{Lu2021EfficientQDS}.
This protocol employs entangled photon pairs to generate virtual $n$-party GHZ states and directly produces secure QSS keys with $O(\eta^{2})$ instead of $O(\eta^{n})$ scaling, where $\eta$ is the transmission rate from a single user to the untrusted central node.

\begin{figure}[t]
\centering
\includegraphics[width=\linewidth]{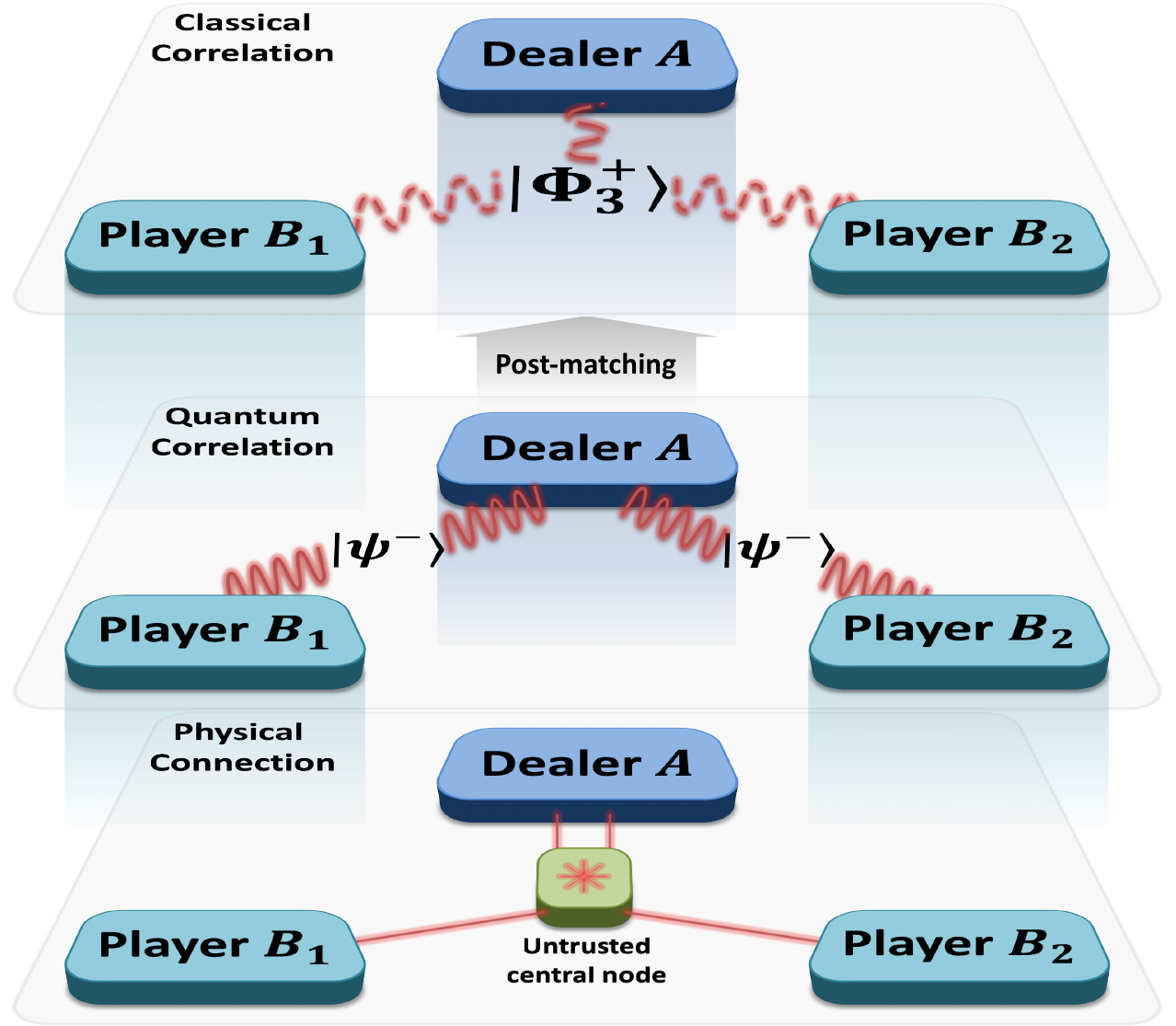}
\caption{
Scheme of the resource-efficient SI QSS protocol.
In the physical connection layer, the untrusted central node distributes the entangled photon pairs between the dealer $A$ and the player $B_1$, and between the dealer $A$ and the player $B_2$, respectively.
In the quantum correlation layer, the dealer $A$ shares the Bell states $\ket{\psi^{-}}$ with each player $B_j$ ($j=1,2$).
With the conduction of the postmatching method, all participants share the equivalent GHZ-state $\ket{\Phi^{+}_{3}} = (\ket{000} + \ket{111}) / \sqrt{2}$ correlations among their classical measurement results.
} 
\label{architecture}
\end{figure}

In this Letter, we utilize two polarization-entangled photon pair sources to conduct the first experimental demonstration of the resource-efficient SI QSS protocol~\cite{xiao2024source} in a three-participant scenario. 
With the finite-key analysis in the composable framework~\cite{Walk2021Sharing}, the QSS keys obtained in our work are secure against any coherent attacks, including attacks by internal participants~\cite{qin2007cryptanalysis}.
Two Sagnac-type entanglement sources~\cite{kim2006phase,fedrizzi2007wavelength,lim2008broadband} are pumped by one ultrafast pulsed laser to generate the entangled photon pairs in the untrusted central node, and then the high-quality photon pairs are distributed to the dealer and each player, as shown in Fig.~\ref{architecture}.
After the postmatching method, the equivalent GHZ-state $\ket{\Phi^{+}_{3}} = (\ket{000} + \ket{111}) / \sqrt{2}$ correlations are established among three participants, yielding the secure QSS key rates of 1.71, 4.69, and 21.18 kbps under single-user quantum channel losses of 12.9, 10.9, and 7.6 dB respectively.
Our experimental results demonstrate the high efficiency and feasibility of the  SI QSS protocol, presenting the potential applications of this protocol in large-scale quantum networks using practical devices to achieve high key rate performance.

\section{Experimental setup}
\begin{figure*}[t]
\centering
\includegraphics[width=\linewidth]{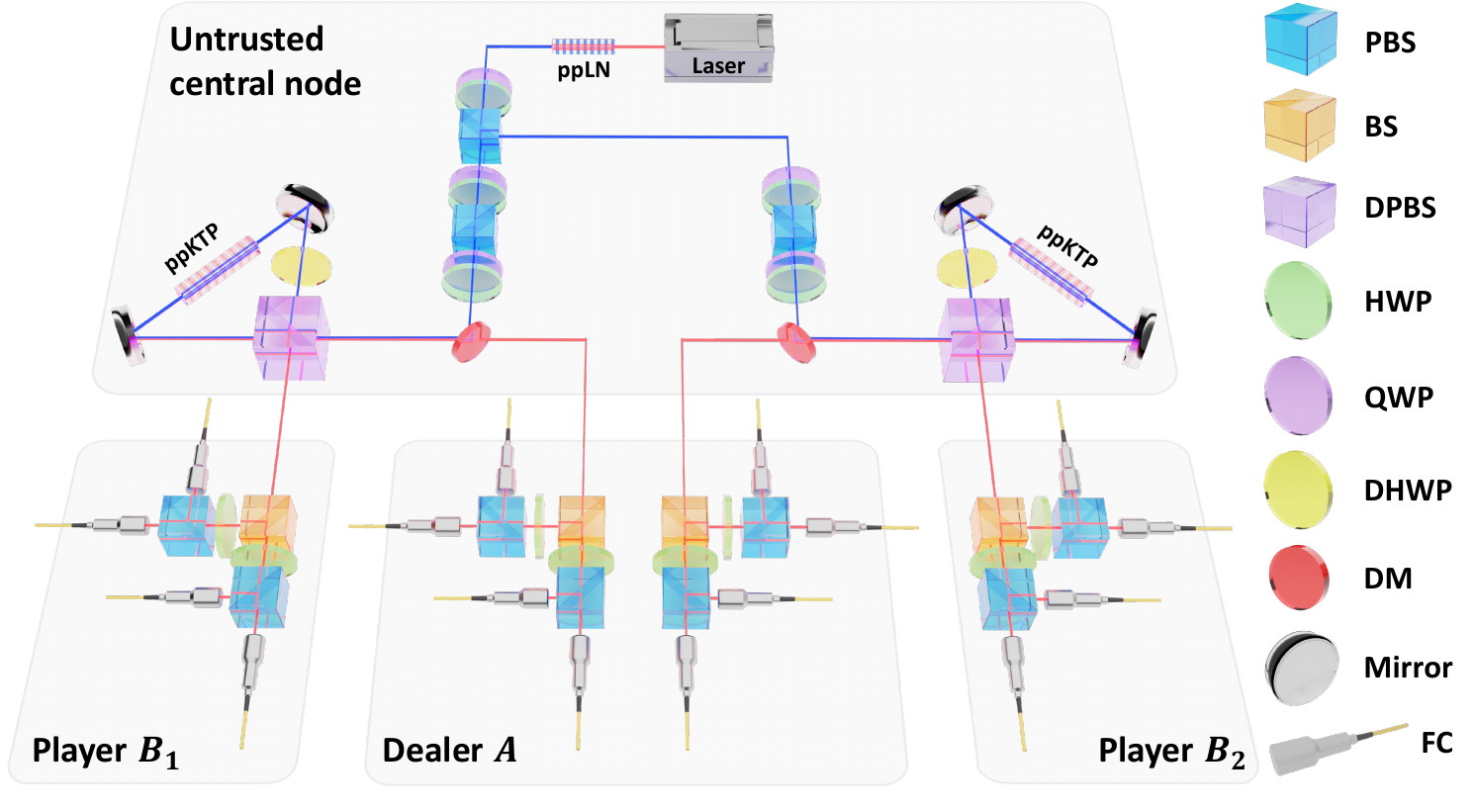}
\caption{
Experimental setup.
The untrusted central node is equipped with two Sagnac-type entanglement sources to distribute Bell states between the dealer $A$ and the player $B_j$ ($j=1,2$).
All protocol participants are equipped with polarization measurement modules to analyze the polarization of received photons.
DPBS, dichroic polarization beam splitter; 
QWP, quarter-wave plate; 
DHWP, dichroic half-wave plate; 
DM, dichroic mirror.
} 
\label{experimental_setup}
\end{figure*}

The physical connection layer contains all the experimental components and presents the physical connection of the resource-efficient SI QSS protocol under our experimental setup, which is shown in Fig.~\ref{experimental_setup}.
The untrusted central node utilizes two polarization-entanglement sources to distribute the Bell states $\ket{\psi^{-}}$ between the dealer $A$ and each player $B_j$ ($j=1,2$). 
Each participant randomly performs rectilinear or diagonal-basis measurement using the polarization measurement module in a passive manner, achieved through beam splitter (BS)~\cite{fung2011universal}.
In the quantum correlation layer, the Bell states $\ket{\psi^{-}}$ are shared between the dealer $A$ and each player $B_j$, respectively. 
After performing the postmatching method and XOR operations, the equivalent GHZ-state correlations can be established among the measurement results of three participants~\cite{xiao2024source}, i.e., the classical correlation layer.
The detailed protocol description is shown in the Appendix~\ref{Description}.

To generate polarization-entangled photon pairs, we adopted a Sagnac-type setup to bidirectionally pump a type-\uppercase\expandafter{\romannumeral 2} periodically poled potassium titanyl phosphate (ppKTP) crystal with a poling period of 46.2 $\mu$m. 
An ultrafast pulsed laser at 1560 nm with a repetition frequency of 96.7 MHz and a pulse duration of 128 fs is frequency doubled in a periodically poled lithium niobate (ppLN) crystal by second-harmonic generation. 
Then the second harmonic is split into two components and launched into two Sagnac-type setups to pump the ppKTP crystals and generate polarization-entangled photon pairs by a spontaneous parametric down-conversion process. 
By adjusting the polarization of the second harmonic, the Sagnac-type setup can generate the polarization-entangled photon pairs with a central wavelength of 1560 nm for both signal and idler photons: 
\begin{equation}
    \ket{\psi^{-}} = \frac{1}{\sqrt{2}} \left(\ket{H}_s\ket{V}_i - \ket{V}_s\ket{H}_i \right),
\end{equation}
where $\ket{H}$ ($\ket{V}$) represents horizontally (vertically) polarized photon, and subscript $s$ ($i$) refers to the signal (idler) photon. 
The original SI QSS protocol~\cite{xiao2024source} utilizes the Bell state $\ket{\phi^{+}} = \left(\ket{HH} + \ket{VV} \right) / \sqrt{2}$ to establish the equivalent GHZ-state correlation among all participants.
However, the Bell state $\ket{\phi^{+}}$ and $\ket{\psi^{-}}$ can be transformed into each other through local unitary operations, which is equivalent to the dealer $A$ flipping the classical bits of both the rectilinear basis and diagonal basis (see details in the Appendix~\ref{Analysis}).
Therefore, our experimental setup is capable of implementing the original protocol.

The generated photon pairs are then coupled into the polarization measurement modules through beam reducers.
Each player $B_j$ is equipped with one polarization measurement module to analyze the polarization of photons, while the dealer $A$ is equipped with two polarization measurement modules to measure the photons corresponding to the different players.
Each polarization measurement module comprises a BS for passively selecting the measurement basis, two half-wave plates (HWPs) and polarization beam splitters (PBSs) for the polarization analysis of photons, and four fiber couplers (FCs) for collecting the photons into optical fibers.
Each FC is connected to one channel of a superconducting nanowire single-photon detector (SNSPD) via optical fibers.
To demonstrate the key rates under different channel losses, we introduce different variable optical attenuators (VOAs) into the fibers between the FCs and the channels of SNSPD.
The entire measurement module requires sixteen detection channels for basis selection and measurement.
However, owing to the postmatching method employed in the resource-efficient SI QSS protocol, the measurements between the dealer $A$ and different players $B_j$ can be performed separately, and then the measurement results from different entanglement sources can be matched through the postmatching method.
As a result, only eight detection channels are needed in our experimental setup.
All the detection channels are linked to a time-to-digital converter to record the timestamps of the detected photons.
Using these timestamps, coincidence counts between different measurement modules are recorded within a temporal window of 5.16 ns, corresponding to the measurements of the entangled photon pairs.

\section{Experimental results}
It is crucial to confirm that the polarization-entanglement sources can generate high-quality entangled photon pairs. 
To verify the quality of the entanglement sources, we first measure the polarization-entanglement visibility in two mutually unbiased bases, i.e., rectilinear basis and diagonal basis.
For idler photons, the fast axis of HWP is oriented at $0^{\circ}$ for the rectilinear basis and $22.5^{\circ}$ for the diagonal basis, and for signal photons, the angle of the HWP fast axis is swept from $0^{\circ}$ to $90^{\circ}$, corresponding to a polarization rotation from $0^{\circ}$ to $180^{\circ}$.
The visibility under the rectilinear basis exceeds $0.990$ for both entanglement sources, while under the diagonal basis, the visibility is $0.984 \pm 0.005$ for source 1 and $0.967 \pm 0.002$ for source 2.
We further measure the fidelity of the generated photon pairs using quantum state tomography~\cite{james2001measurement}, which involves projecting both signal and idler photons onto four polarization states $\{\ket{H}, \ket{V}, \ket{D}, \ket{R}\}$ for coincidence counts, where $\ket{D}$ and $\ket{R}$ represent the diagonally polarized photon and right-circularly polarized photon, respectively.
The fidelity of the entangled photon pairs with respect to the Bell state $\ket{\psi^-}$ is $0.988 \pm 0.001$ for source 1 and $0.987 \pm 0.001$ for source 2.
Table~\ref{Source_data} shows the main parameters of two polarization-entanglement sources, which presents the high quality of the entangled photon pairs generated by two sources.

\begin{table}[h!]
\renewcommand{\arraystretch}{1.2}
\centering
\caption{Main parameters of the polarization-entanglement source. 
$V_{\mathrm{rec}}$ and $V_{\mathrm{dia}}$ represent the visibility under the rectilinear basis and the diagonal basis, respectively. 
$F$ is the fidelity of the entangled photon pairs. 
$\mu$ represents the average number of photon pairs generated by one pulse.
}
\begin{tabular*}{0.48\textwidth}{@{\extracolsep{\fill}}cccc}
\hline
Source & $V_{\mathrm{rec}}$ & $V_{\mathrm{dia}}$ & $F$ \\
\hline
Source 1 & 0.995 & 0.984 & 0.988 \\
($\mu = 0.004$) & ($\pm 0.001$) & ($\pm 0.005$) & ($\pm 0.001$) \\
\hline
Source 2 & 0.991 & 0.967 & 0.987 \\
($\mu = 0.002$) & ($\pm 0.001$) & ($\pm 0.002$) & ($\pm 0.001$) \\
\hline
\end{tabular*}
\label{Source_data}
\end{table}

\begin{table*}[htbp]
\renewcommand{\arraystretch}{1.25}
\centering
\caption{Summary of experimental data for the source-independent QSS protocol.
The key rates are tested with different channel losses and probabilities of basis selection. 
The channel loss is the average loss between the untrusted central node and the participants.
$N$ denotes the total number of pulses.
$R_{\mathrm{QSS}}$ and $r_{\mathrm{QSS}}$ denote the experimental QSS key rate.}
\begin{tabular*}{\textwidth}{@{\extracolsep{\fill}}cccccccccc}
\hline
Channel loss (dB) & $p_x$ & $N$ & $E_{AB_1}^X$ & $E_{AB_2}^X$ & $E^X$ & $E_{AB_1}^Z$ & $E_{AB_2}^Z$ & $R_{\mathrm{QSS}}$ (per pulse) & $r_{\mathrm{QSS}}$ (kbps) \\ 
\hline
7.6 & 0.9 & $10^{11}$ & 0.019 & 0.017 & 0.035 & 0.020 & 0.026 & $2.19\times10^{-4}$ & $21.177$ \\
10.9 & 0.9 & $10^{11}$ & 0.021 & 0.016 & 0.036 & 0.022 & 0.021 & $4.85\times10^{-5}$ & $4.689$ \\
12.9 & 0.9 & $10^{11}$ & 0.021 & 0.018 & 0.038 & 0.025 & 0.023 & $1.77\times10^{-5}$ & $1.711$ \\
\hline
7.8 & 0.5 & $10^{11}$ & 0.018 & 0.020 & 0.037 & 0.020 & 0.019 & $6.28\times10^{-5}$ & $6.072$ \\
10.7 & 0.5 & $10^{11}$ & 0.020 & 0.021 & 0.040 & 0.021 & 0.021 & $1.56\times10^{-5}$ & $1.508$ \\
12.6 & 0.5 & $10^{11}$ & 0.020 & 0.021 & 0.040 & 0.021 & 0.021 & $6.44\times10^{-6}$ & $0.622$ \\
\hline
\end{tabular*}
\label{experimental_results_1}
\end{table*}

To achieve a lower quantum bit error rate (QBER) in our experimental demonstration, the rectilinear basis is regarded as the $X$ basis and the diagonal basis is regarded as the $Z$ basis, which is equivalent to the original protocol.
The average number of photon pairs per pulse is increased to $\mu = 0.023$ for source 1 and $\mu = 0.021$ for source 2 to achieve a higher key rate (the detailed performances are shown in the Appendix~\ref{performance}).
We use BS with distinct split ratios to implement different probabilities of basis selection, $R:T = 50:50$ for the balanced selection ($p_x = 0.5$) and $R:T = 90:10$ for the biased selection ($p_x = 0.9$), where $R$ and $T$ represent the reflectivity and transmittance of the BS, and $p_x$ denotes the selection probability of the $X$ basis.
Under both the balanced and biased selection, we perform the experiments with three average single-user channel losses, i.e., the channel loss between the untrusted central node and one participant, which incorporate both the collection efficiency of FCs and the attenuation of VOAs between FCs and channels of SNSPD.
According to the resource-efficient SI QSS protocol~\cite{xiao2024source}, the length of secure QSS key is 
\begin{equation}
l = n^{X} \left[1 - \max_{j} h\left(\bar{\phi}^X_{AB_j}\right) - f_e h\left(E^X\right)\right] - \log _{2} \frac{1}{4 \epsilon_{\mathrm{c}} \left( \epsilon^\prime \right)^2},
\label{QSS key length}
\end{equation}
where $\bar{\phi}^X_{AB_j}$ denotes the $X$-basis quantum phase error rate between the dealer $A$ and the player $B_j$, which is estimated from the $Z$-basis QBER $E^Z_{AB_j}$ using the method of random sampling without replacement~\cite{yin2020tight}.
$E^X$ represents the $X$-basis QBER among all participants.
Table~\ref{experimental_results_1} presents all the experimental $X$- and $Z$-base QBERs under the different basis selection probabilities and average channel losses.
$n^{X}$ denotes the coincidence counts under the $X$ basis.
$h(x)=-x\log_{2}(x)-(1-x)\log_{2}(1-x)$ is the binary Shannon entropy function.
The error correction inefficiency $f_e = 1.16$, and the failure probabilities $\epsilon^\prime = \epsilon_{\mathrm{c}} = 10^{-10}$ for both theoretical simulation and experimental demonstration.
See the Appendix~\ref{Analysis} for the detailed parameters and security analysis.

\begin{figure}[t]
\centering
\includegraphics[width=\linewidth]{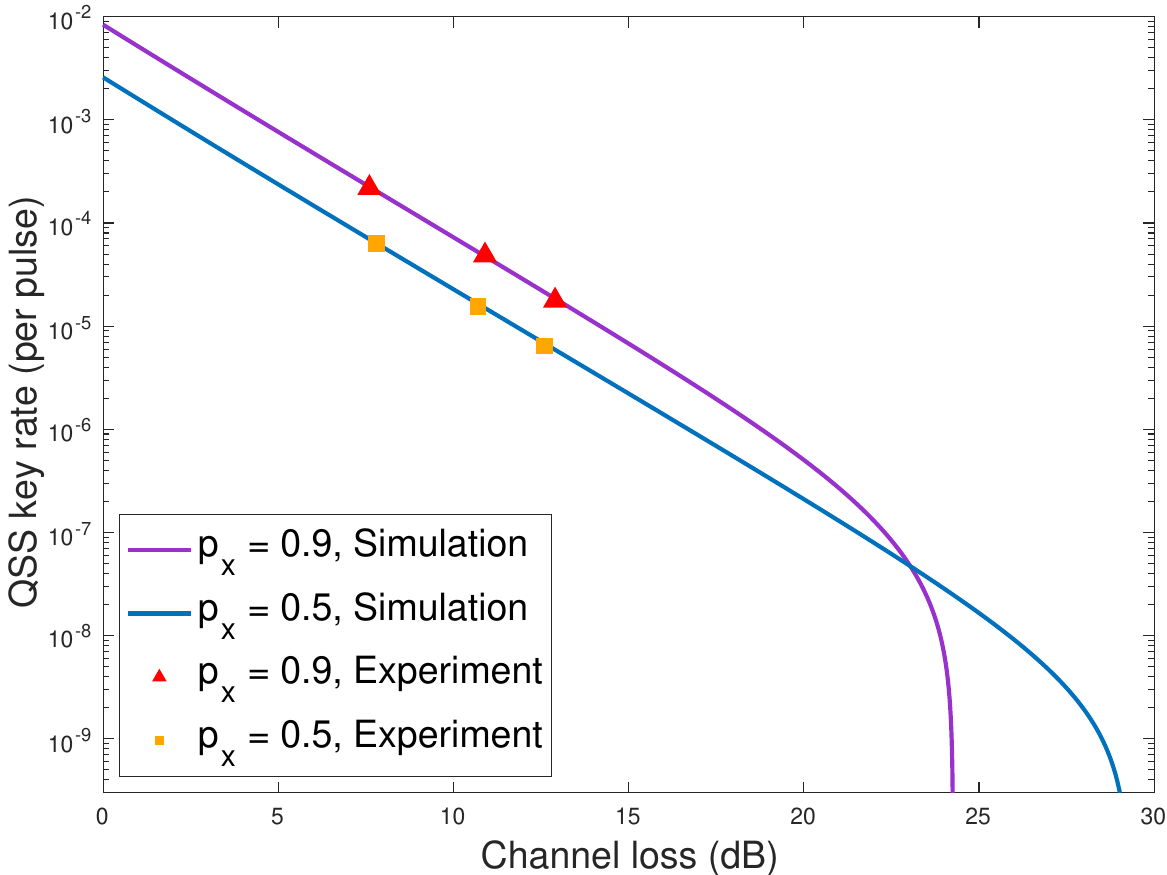}
\caption{
Experimental results of the resource-efficient SI QSS.
We plot the experimental data in triangle scatters for $p_x = 0.9$ and square scatters for $p_x = 0.5$.
Channel loss represents the average loss between the untrusted central node (entanglement sources) and the participants (detection channels of SNSPD).
} 
\label{QSS_keyrate}
\end{figure}

Figure~\ref{QSS_keyrate} shows the experimental key rates and theoretical simulation of the resource-efficient SI QSS protocol.
In theoretical simulation, we set the misalignment error rate $e_d = 0.01$, the detection efficiency $\eta_d = 0.83$, and the dark count rate (per pulse) $p_d = 1.3 \times 10^{-7}$ based on the average parameters of eight channels (see the detailed channel parameters in the Appendix~\ref{performance}).
With the pump repetition frequency of 96.7 MHz, the highest experimental SI QSS key rate under the biased selection could achieve 21.18 kbps with an average channel loss of 7.6 dB between the untrusted central node and each participant.
Compared to the previous QSS protocol using multipartite bound entanglement~\cite{Zhou2018BoundEntangleQSS}, which achieved a QSS key rate of 2.25 kbps, our experimental results demonstrate an improvement in efficiency, highlighting the potential of the resource-efficient SI QSS protocol in future large-scale quantum networks.
As the average single-user channel loss increases to 10.9 dB and 12.9 dB, the experimental QSS key rates remain at 4.689 kbps and 1.711 kbps, respectively.
Under the balanced selection, the experimental results of key rates could also achieve 6.072, 1.508, and 0.622 kbps with different single-user channel losses, as shown in Table~\ref{experimental_results_1}.
The experimental results with different probabilities of basis selection show that parameter optimization will significantly affect the secure key rate, which will be particularly important in networks with more users.

Furthermore, we present a proof-of-principle experimental demonstration of the scalability.
Here, we implement this by using a single entangled source to distribute entangled photon pairs and performing repeated measurements to obtain multiple datasets.
Different datasets correspond to the measurement results between dealer $A$ and different players $B_i$.
Through the postmatching method, the measurement results in different datasets can be correlated, making it possible to simulate the distribution of virtual $n$-party GHZ states among multiple users.
All measurement results are obtained under balanced basis selection and identical channel loss (7.6 dB) conditions.
Table~\ref{experimental_results_2} shows the experimental key rates with 6.972, 6.461, and 5.880 kbps for the scenarios with three, four, and five participants, respectively.
The result indicates that the experimental key rate decreases only slightly as the number of participants increases, suggesting that the resource-efficient SI QSS protocol exhibits scalability (See more details in the Appendix~\ref{scalability and robustness}).

\begin{table}[htbp]
\renewcommand{\arraystretch}{1.3}
\centering
\caption{Summary of experimental data for the source-independent QSS protocol with different numbers of users.}
\begin{tabular*}{0.48\textwidth}{@{\extracolsep{\fill}}cccccc}
\hline
$n$ & $p_x$ & $N$ & $E^X$ & $\max_{j} \bar{\phi}^X_{AB_j}$ & $r_{\mathrm{QSS}}$ (kbps) \\ 
\hline
3 & 0.5 & $10^{11}$ & 0.020 & 0.027 & $6.972$ \\
4 & 0.5 & $10^{11}$ & 0.030 & 0.027 & $6.461$ \\
5 & 0.5 & $10^{11}$ & 0.040 & 0.027 & $5.880$ \\
\hline
\end{tabular*}
\label{experimental_results_2}
\end{table}

\section{Discussion}
We have experimentally demonstrated the resource-efficient SI QSS protocol by utilizing two high-quality Sagnac-type entangled photon pair sources.
By employing the postmatching method~\cite{Lu2021EfficientQDS}, the SI QSS key rate in the three-participant scenario can be improved from $\mathcal{O}\left(\eta^{3} \right)$ to $\mathcal{O}\left(\eta^{2} \right)$, which enables our experiment to achieve a high key rate of 21.18 kbps with an average single-user channel loss of 7.6 dB, demonstrating an improvement in efficiency compared to the previous protocols.
We also demonstrate the scalability of this protocol by employing a single entanglement source with measurement results in different datasets to generate secure keys for cases with three, four, and five participants.
Additionally, there is no requirement for the synchronization and coincident measurements among all participants in our experimental demonstration, which demonstrates great experimental feasibility of the resource-efficient SI QSS protocol.
Despite these notable results achieved in our experimental demonstration, it is worth noting that collection efficiency limits our experimental key rates. 
These key rates could be increased by at least 1 order of magnitude if our experimental system could reach the efficiency of device-independent implementations~\cite{liu2021device}.
Besides, the significant advantage of this protocol in scalability can be particularly demonstrated in the fully connected quantum networks~\cite{Wengerowsky2018EntangleNetwork,Joshi2020EntangleNetwork,liu202240,huang2025sixteen}, where it enables real-time and high-efficient QSS among multiple users.

\section{Acknowledgments}
This work is supported by the National Natural Science Foundation of China (No. 12274223) and the Fundamental Research Funds for the Central Universities and the Research Funds of Renmin University of China (No. 24XNKJ14).

\appendix
\section{Source-independent quantum secret sharing protocol}
\label{SI QSS}

\subsection{Protocol description}
\label{Description}
Here, we present a brief description of the source-independent quantum secret sharing (SI QSS) protocol~\cite{xiao2024source} in a multi-participant scenario, which can be readily simplified to the case involving three participants.

\textbf{Distribution and measurement.} 
The untrusted central node prepares entangled photon pairs in the state $\ket{\phi^{+}}$, which are distributed among $n$ participants.
One photon from each photon pair is sent to the dealer $A$, while the other photon is distributed to a player $B_j$ among all $n-1$ players, ensuring that the dealer $A$ and each player $B_j$ could share entangled photon pairs.
Then all participants perform measurements on their own photons in the $X$-basis with probability $p_x$ and the $Z$-basis with probability $1-p_x$.

In the original SI QSS protocol~\cite{xiao2024source}, distribution is achieved through the fully connected quantum network~\cite{Wengerowsky2018EntangleNetwork,Joshi2020EntangleNetwork,liu202240,huang2025sixteen}, including a single entanglement source and wavelength-division multiplexing.
In our experimental demonstration, we employ two entanglement sources to implement an equivalent experimental setup for the three-participant scenario.

\textbf{postmatching.} 
The basis choices and timestamps of all measurement events are broadcast through the classical channels.
The coincidental measurement events that the dealer $A$ and one player $B_j$ both select the same basis are retained as matching events.
$a_{j}^X$ ($a_{j}^Z$) represents the $X$-basis ($Z$-basis) measurement results of matching events for the dealer $A$ and $b_{j}^X$ ($b_{j}^Z$) is the corresponding $X$-basis ($Z$-basis) measurement results for the player $B_j$.

Then all participants perform classical operations on their classical bits as follows.
The dealer $A$ preserves the classical bits $\tilde{a}_{1}^X = a_{1}^X \oplus a_{2}^X \oplus \cdots \oplus a_{n-1}^X$ and broadcasts the value of $a_{1}^Z \oplus a_{j}^Z$.
Each player $B_j$ flips the classical bit $b_{j}^Z$ and records it as $\tilde{b}_{j}^Z$ according to the corresponding value of $a_{1}^Z \oplus a_{j}^Z$, i.e., if $a_{1}^Z \oplus a_{j}^Z = 1$, $b_{j}^Z$ is flipped, otherwise, no operations are performed on $b_{j}^Z$.
After the above operations, the classical bits of all participants satisfy the correlations: 
\begin{equation}
\begin{split}
	\tilde{a}_{1}^X = b_{1}^X \oplus b_{2}^X \oplus \cdots \oplus b_{n-1}^X ,\\
	a_{1}^Z = \tilde{b}_{1}^Z = \tilde{b}_{2}^Z = \cdots = \tilde{b}_{n-1}^Z ,
\end{split}
\end{equation}
which are equivalent to the GHZ states shared among all participants.

\textbf{Parameter estimation and post processing.}
After above processing, the $X$-basis classical bits $\{\tilde{a}_{1}^X, b_{1}^X, b_{2}^X, \cdots,\\ b_{n-1}^X\}$ are preserved for key generation and the $Z$-basis classical bits $\{a_{1}^Z, \tilde{b}_{1}^Z, \tilde{b}_{2}^Z, \cdots, \tilde{b}_{n-1}^Z\}$ are preserved for parameter estimation.
To conduct parameter estimation, all participants disclose the $Z$-basis classical bits in any order.
Then the dealer $A$ selects a random subset of the $Z$-basis classical bits and calculates the QBER with each single player $B_j$.
The largest $Z$-basis QBER is kept as the results of parameter estimation.

After all participants obtain the correlated raw keys, they proceed with the error correction and obtain the corrected keys.
Then they conduct the privacy amplification according the results of parameter estimation, which will extract $l$ bits final keys.

The length of final key is given by~\cite{xiao2024source}:
\begin{equation}
l = n^{X} \left[1 - \max_{j} h\left(\bar{\phi}^X_{AB_j}\right) - f_e h\left(E^X\right)\right] - \log _{2} \frac{1}{4 \epsilon_{\mathrm{c}} \left( \epsilon^\prime \right)^2},
\label{finite key length}
\end{equation}
where $\bar{\phi}^X_{AB_j} = E^Z_{AB_j} + \gamma\left(E^Z_{AB_j},\bar{\epsilon} \right)$ denotes the $X$-basis quantum phase error rate between the dealer $A$ and the player $B_j$, which is estimated from the $Z$-basis QBER $E^Z_{AB_j}$ using the method of random sampling without replacement~\cite{yin2020tight}.
In this method, we have $\gamma\left(\lambda,\bar{\epsilon}\right) = \frac{\frac{\left(1 - 2 \lambda \right)AG}{m+k} + \sqrt{\frac{A^2G^2}{\left(m+k\right)^2} + 4\lambda\left(1-\lambda\right)G}}{2 + 2\frac{A^2G}{\left(m+k\right)^2}}$, 
where $\lambda = E^{Z}_{AB_j}$, $A = \max\{m, k\}$, $G = \frac{m+k}{mk} \ln \frac{m+k}{2\pi mk\lambda \left(1-\lambda\right) \bar{\epsilon}^2}$,
$m = n^{X}$ denotes the length of $X$-basis classical bits as the raw keys and $k = n^{Z}$ denotes the length of $Z$-basis classical bits for the parameter estimation between the dealer $A$ and the player $B_j$.
$E^X$ represents the $X$-basis QBER among all participants and $f_e=1.16$ represents the error correction efficiency.
$h(x)=-x\log_{2}(x)-(1-x)\log_{2}(1-x)$ is the binary Shannon entropy function.
The failure probabilities are set to $\bar{\epsilon} = \epsilon^\prime = \epsilon_{\mathrm{c}} = 10^{-10}$.

\subsection{Security analysis}
\label{Analysis}
The complete security analysis has been provided in the original SI QSS protocol~\cite{xiao2024source}.
Therefore, here we give a supplementary security analysis of our experimental demonstration and a brief analysis for the participant attacks~\cite{karlsson1999quantum,qin2007cryptanalysis}.

\begin{figure}[t]
\centering
\includegraphics[width=\linewidth]{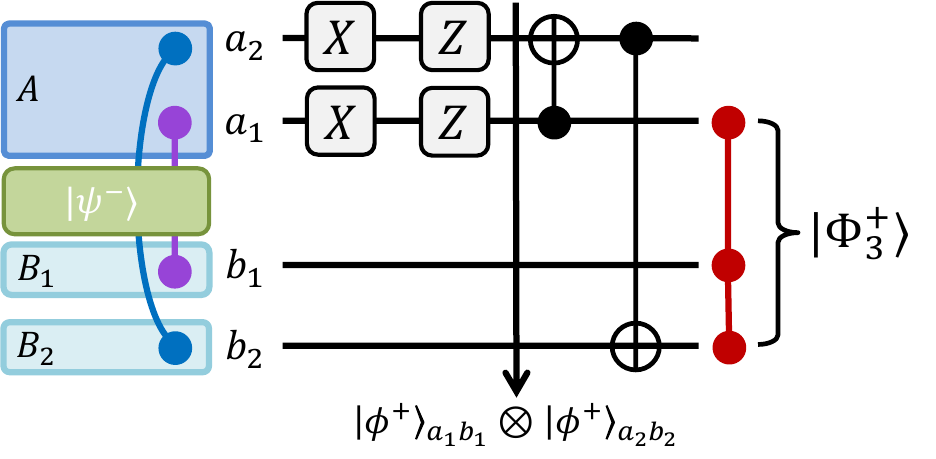}
\caption{
Schematic for the generation of GHZ state through quantum operations in three-participant scenario, which is equivalent to our experimental demonstration.
The Bell state $\ket{\psi^-}$ is transformed into $\ket{\phi^+}$ through the local unitary operations $Z$ gate and $X$ gate.
Then CNOT gates are performed on two Bell states $\ket{\phi^{+}}_{a_1b_1} \otimes \ket{\phi^{+}}_{a_2b_2}$ and the three-party GHZ state $\ket{\Phi^+_3}$ is established among $a_1$, $b_1$ and $b_2$.
} 
\label{qcircuit}
\end{figure}

In the original protocol, the security is proved by introducing an equivalent virtual protocol, where the GHZ states are established among all participants through quantum memories and quantum operations.
In our experimental demonstration, the entanglement sources generate the Bell state $\ket{\psi^-}$, which is distinct with the Bell state $\ket{\phi^+}$ utilized in the original protocol.
However, the Bell states above can be transformed into each other through the local unitary operations, i.e., 
\begin{equation}
\begin{split}
    \ket{\psi^-} \xrightarrow[\mathrm{Unitary}]{\mathrm{Local}} 
    & \left(Z \otimes I\right)\left(X \otimes I\right)\ket{\psi^-} \\
    = & \frac{1}{\sqrt{2}} \left(Z X \ket{1}\ket{0} - Z X \ket{0}\ket{1} \right) \\
    = & \frac{1}{\sqrt{2}} \left(\ket{0}\ket{0} + \ket{1}\ket{1} \right) \\
    = & \ket{\phi^+}.
\end{split}
\label{bell_state_transform}
\end{equation}

Therefore, in the three-participant scenario, the quantum operations in the virtual protocol of the original security analysis, which are used to establish the GHZ states among all participants, can be adapted into the quantum circuit shown in Fig.~\ref{qcircuit}.
The untrusted central node distributes the Bell state $\ket{\psi^-}_{a_1b_1}$ between the dealer $A$ and the player $B_1$, and the Bell state $\ket{\psi^-}_{a_2b_2}$ between the dealer $A$ and the player $B_2$.
The dealer $A$ performs $X$ gates and $Z$ gates on the received virtual qubits $a_1$ and $a_2$. 
According to the equation~\ref{bell_state_transform}, the state $\ket{\psi^-}_{a_1b_1} \otimes \ket{\psi^-}_{a_2b_2}$ is transformed into the state $\ket{\phi^{+}}_{a_1b_1} \otimes \ket{\phi^{+}}_{a_2b_2}$, which is consistent with the case in the original protocol.
All participants then generate the almost perfect GHZ states following the steps in the original virtual protocol.

In our experimental demonstration, after measuring the received photons, the dealer $A$ flips the classical bits in rectilinear basis ($X$-basis), which is equivalent to the $Z$ gate performed on the virtual qubits.
The dealer $A$ also flips the classical bits in diagonal basis ($Z$-basis), which is equivalent to the $X$ gate.
Then all participants generate the final secure keys according to the original actual protocol.
Therefore, the actual experimental demonstration is equivalent to the alternate virtual protocol, where the almost perfect GHZ states are shared among all participants.
The security is guaranteed by the monogamy of entanglement~\cite{Terhal2004EntangleMonogamous}, and the information exposed to eavesdroppers is negligible, allowing all participants to obtain the information-theoretically secure keys.

For the attacks from internal eavesdroppers, which are known as participant attacks, the SI QSS protocol utilizes the method in refs.~\cite{Kogias2017CVQSS,Walk2021Sharing}, where the dealer $A$ calculates the $Z$-basis QBER with each single player $B_j$ to defend against the internal malicious participants.
The original protocol is an ($n$-$n$)-threshold QSS scheme, so the secret of the dealer can only be reconstructed by the cooperation of all $n$ players, which requires that any collaboration of $n-1$ players (unauthorized subset) should never be able to decode the secret.
Therefore, any unauthorized subset $U_j$ ($j=1,2,\cdots,n$) should be regarded as the eavesdropper during the process of parameter estimation.
For each unauthorized subset $U_j$, the parameter estimation could be simplified to a two-party scenario, involving the dealer $A$ and the remaining player $B_j$.
They calculate the $Z$-basis QBER with each other to eliminate the influence of the players in unauthorized subset $U_j$.
Then the dealer $A$ selects the maximum $Z$-basis QBER as the result of parameter estimation.
Following the ref.~\cite{Walk2021Sharing}, the composable security of the original protocol with the above method against the participant attacks is presented below.

Through the postmatching method, all participants obtain the correlated raw keys in $X$-basis, i.e., $\mathbf{X}_{A}$ for the dealer $A$ and $\mathbf{X}_{B_{j}}$ for the player $B_{j}$.
Then all participants will extract the final keys $\mathbf{S}_{A}$ and $\mathbf{S}_{B_{j}}$ with post processing.
In the composable security, the attacks of adversary can involve any operations permitted by quantum laws rather than being restricted to specific attacks.
In general, the final key $\mathbf{S}_{A}$ of dealer $A$ is quantum mechanically correlated with a quantum state that is held by the adversary, which can be described by a classical-quantum state
\begin{equation}
    \rho_{\mathbf{S}_{A}, EU_{j}}=\sum_{\mathbf{S}_{A}} p(\mathbf{S}_{A}) \ket{\mathbf{S}_{A}}\bra{\mathbf{S}_{A}} \otimes \rho_{EU_{j}}^{\mathbf{S}_{A}},
\label{C-Q_state}
\end{equation}
where the sum is taken over all possible strings of $\mathbf{S}_{A}$ and $p(\mathbf{S}_{A})$ is the probability of each possible string, $\rho_{EU_{j}}^{\mathbf{S}_{A}}$ is the joint state of the eavesdropper $E$ and the $j$th unauthorized subset $U_{j}$ given a certain string of final key $\mathbf{S}_{A}$.
Ideally, a QSS protocol is secure if it satisfies the correctness and the secrecy.
The correctness requires that the final key of dealer $A$ is identical to the string reconstructed by all players, i.e., $\mathbf{S}_{A} = \mathbf{S}_{\mathrm{players}}$.
The secrecy requires that the joint state of the eavesdropper $E$ and the $j$th unauthorized subset $U_{j}$ is completely decoupled from the dealer $A$, i.e., $\rho_{\mathbf{S}_{A}, EU_{j}} = \sum_{\mathbf{S}_{A}} \frac{1}{|\mathbf{S}_{A}|} \ket{\mathbf{S}_{A}}\bra{\mathbf{S}_{A}} \otimes \sigma_{EU_{j}}$, which means that the system of dealer $A$ is the uniform mixture of all possible strings $\mathbf{S}_{A}$.

However, these two conditions cannot be perfectly met in practical situations, which implies that the conditions of security should tolerate some minuscule errors.
A QSS protocol is $\epsilon_{\mathrm{c}}$-correct if 
\begin{equation}
    p\left(\mathbf{S}_{A} \neq \mathbf{S}_{\mathrm{players}}\right) \leq \epsilon_{\mathrm{c}},
\end{equation}
which means that the probability that the final key $\mathbf{S}_{A}$ and the reconstructed string $\mathbf{S}_{\mathrm{players}}$ are not identical is not greater than $\epsilon_{\mathrm{c}}$.
A QSS protocol is $\epsilon_{\mathrm{s}}$-secret if 
\begin{equation}
    \max_{j}\left\{p_{\mathrm{pass}} D\left( \rho_{\mathbf{S}_{A}, EU_{j}}, \sum_{\mathbf{S}_{A}} \frac{1}{|\mathbf{S}_{A}|}\ket{\mathbf{S}_{A}}\bra{\mathbf{S}_{A}} \otimes \sigma_{EU_{j}}\right)\right\} \leq \epsilon_{\mathrm{s}},
\end{equation}
which means that the maximum distance $D(\cdot,\cdot)$ between the classical-quantum
state $\rho_{\mathbf{S}_{A}, EU_{j}}$ and the decoupled state $\sum_{\mathbf{S}_{A}} \frac{1}{|\mathbf{S}_{A}|} \ket{\mathbf{S}_{A}}\bra{\mathbf{S}_{A}} \otimes \sigma_{EU_{j}}$ is not greater than $\epsilon_{\mathrm{s}} / p_{\mathrm{pass}}$, where $p_{\mathrm{pass}}$ is the probability that the QSS protocol is not aborted.
If a QSS protocol satisfies two conditions above, it is called $\epsilon_{\mathrm{sec}}$-secure, where $\epsilon_{\mathrm{sec}} \geq \epsilon_{\mathrm{c}} + \epsilon_{\mathrm{s}}$.

In the original source-independent QSS protocol proposed in ref.~\cite{xiao2024source}, after the error correction, all participants compute and compare a hash of length $\log_{2}(1/\epsilon_{\mathrm{c}})$ bits by employing a random universal$_2$ hash function on the raw keys $\mathbf{X}_{A}$ and $\mathbf{X}_{\mathrm{players}}$.
Owing to the property of universal$_2$ hash function~\cite{tomamichel2012tight}, where the probability that the two hash values coincide if $\mathbf{X}_{A}$ and $\mathbf{X}_{\mathrm{players}}$ are different is not greater than $2^{\lceil \log_{2} \epsilon_{\mathrm{c}} \rceil} \leq \epsilon_{\mathrm{c}}$, the $\epsilon_{\mathrm{c}}$-correctness can be satisfied, i.e., $p\left(\mathbf{S}_{A} \neq \mathbf{S}_{\mathrm{players}}\right) \leq p\left(\mathbf{X}_{A} \neq \mathbf{X}_{\mathrm{players}}\right) \leq \epsilon_{\mathrm{c}}$. 

For the $\epsilon_{\mathrm{s}}$-secrecy of the original QSS protocol, we first introduce the quantum leftover hashing Lemma~\cite{renner2008security,tomamichel2011leftover}.
Given a raw key $\mathbf{X}$, a random hash function maps it to a final key $\mathbf{S}$ of length $l$, then for any $\epsilon > 0$, the distance between the classical-quantum state and the decoupled state can be bounded as 
\begin{equation}
    D\left(\rho_{\mathbf{S}, E}, \sum_{\mathbf{S}} \frac{1}{|\mathbf{S}|}\ket{\mathbf{S}}\bra{\mathbf{S}} \otimes \sigma_{E}\right) \leq \sqrt{2^{l-H_{\min}^{\epsilon}\left(\mathbf{X} | E^{\prime}\right)-2}} + \epsilon,
\end{equation}
where $E$ represents the finite or infinite dimensional system of adversary (the eavesdropper and the unauthorized subset) and $E^{\prime}$ represents all information adversary obtained, including the classical communication during the post processing.
Combining the quantum leftover hashing Lemma and the condition of $\epsilon_{\mathrm{s}}$-secrecy, the length of final key satisfies
\begin{equation}
    l \leq \min_{j} H_{\min}^{\epsilon}\left(\mathbf{X}_{A} | E U_{j}^{\prime} \right) - 2\log_{2} \frac{1}{\epsilon^{\prime}} + 2,
\end{equation}
where we choose $\epsilon^{\prime} = \epsilon_{\mathrm{s}} - p_{\mathrm{pass}} \epsilon > 0$ and use the fact that $\log_{2} p_{\mathrm{pass}} < 0$.
During the post processing, $\mathrm{leak}_{\mathrm{EC}} + \log_{2}(1/\epsilon_{\mathrm{c}})$ bits of information about the raw key $\mathbf{X}_{A}$ are leaked through classical channel, so we have~\cite{tomamichel2012tight}
\begin{equation}
    H_{\min}^{\epsilon}\left(\mathbf{X}_{A} | E U_{j}^{\prime}\right) \geq H_{\min}^{\epsilon}\left(\mathbf{X}_{A} | E U_{j}\right) - \mathrm{leak}_{\mathrm{EC}} - \log_{2} \frac{1}{\epsilon_{\mathrm{c}}},
\end{equation}
and the length of final key satisfies
\begin{equation}
    l \leq \min_{j} H_{\min}^{\epsilon}\left(\mathbf{X}_{A} | E U_{j}\right) - \mathrm{leak}_{\mathrm{EC}} - \log_{2} \frac{1}{4 \epsilon_{\mathrm{c}} \left(\epsilon^\prime\right)^{2}}.
\end{equation}
The conditional smooth min-entropy $H_{\min}^{\epsilon}\left(\mathbf{X}_{A} | E U_{j}\right)$ can be bounded through the entropic uncertainty relation~\cite{tomamichel2011uncertainty} 
\begin{equation}
    H_{\min}^{\epsilon}\left(\mathbf{X}_{A} | E U_{j}\right) + H_{\max}^{\epsilon}\left(\mathbf{Z}_{A} | B_{j}\right) \geq n^{X} q,
\end{equation}
where $n^{X}$ is the length of raw key $\mathbf{X}_{A}$, $q$ quantifies the complementarity of the two measurement bases, $B_{j}$ denotes the remaining single player corresponding to the $j$th unauthorized subset $U_{j}$.
The conditional smooth max-entropy $H_{\max}^{\epsilon}\left(\mathbf{Z}_{A} | B_{j}\right)$ represents the correlations between the $Z$-basis raw keys of the dealer $A$ and the player $B_{j}$, which are counterfactual as all qubits for key generation are measured in $X$-basis.
Therefore, the correlations should be estimated via the $Z$-basis classical bits for the parameter estimation.
According to the statistical fluctuation analysis of random sampling without replacement~\cite{tomamichel2012tight,yin2020tight}, we have 
\begin{equation}
    H_{\max}^{\epsilon}\left(\mathbf{Z}_{A} | B_{j}\right) \leq n^{X} h\left(E^Z_{AB_j} + \gamma\left(E^Z_{AB_j}, \bar{\epsilon}\right)\right),
\end{equation}
where $\gamma\left(\lambda,\bar{\epsilon}\right)$ is presented in section~\ref{Description} and $\epsilon = \sqrt{\bar{\epsilon}/p_{\mathrm{pass}}}$.
Finally, we could obtain the length $l$ of finite key as
\begin{equation}
\begin{split}
    l = & n^{X} \left[q - \max_{j} h\left(E^Z_{AB_j} + \gamma\left(E^Z_{AB_j}, \bar{\epsilon}\right)\right)\right] \\
    & - \mathrm{leak}_{\mathrm{EC}} - \log _{2} \frac{1}{4 \epsilon_{\mathrm{c}} \left( \epsilon^\prime \right)^2} ,
\end{split}
\end{equation}
where $q$ is set to 1 in our experimental demonstration and $\mathrm{leak}_{\mathrm{EC}} = n^{X} f_e h\left(E^X\right)$ represents the information leakage of error correction.
Substituting these values and equations, the length of finite key can be rewritten as 
\begin{equation}
l = n^{X} \left[1 - \max_{j} h\left(\bar{\phi}^X_{AB_j}\right) - f_e h\left(E^X\right)\right] - \log _{2} \frac{1}{4 \epsilon_{\mathrm{c}} \left( \epsilon^\prime \right)^2},
\end{equation}
which is the same as equation~\ref{finite key length}.
Therefore, the $\epsilon_{\mathrm{s}}$-secrecy can also be satisfied for the original SI QSS protocol.

\begin{figure*}[t]
\centering
\includegraphics[width=.9\linewidth]{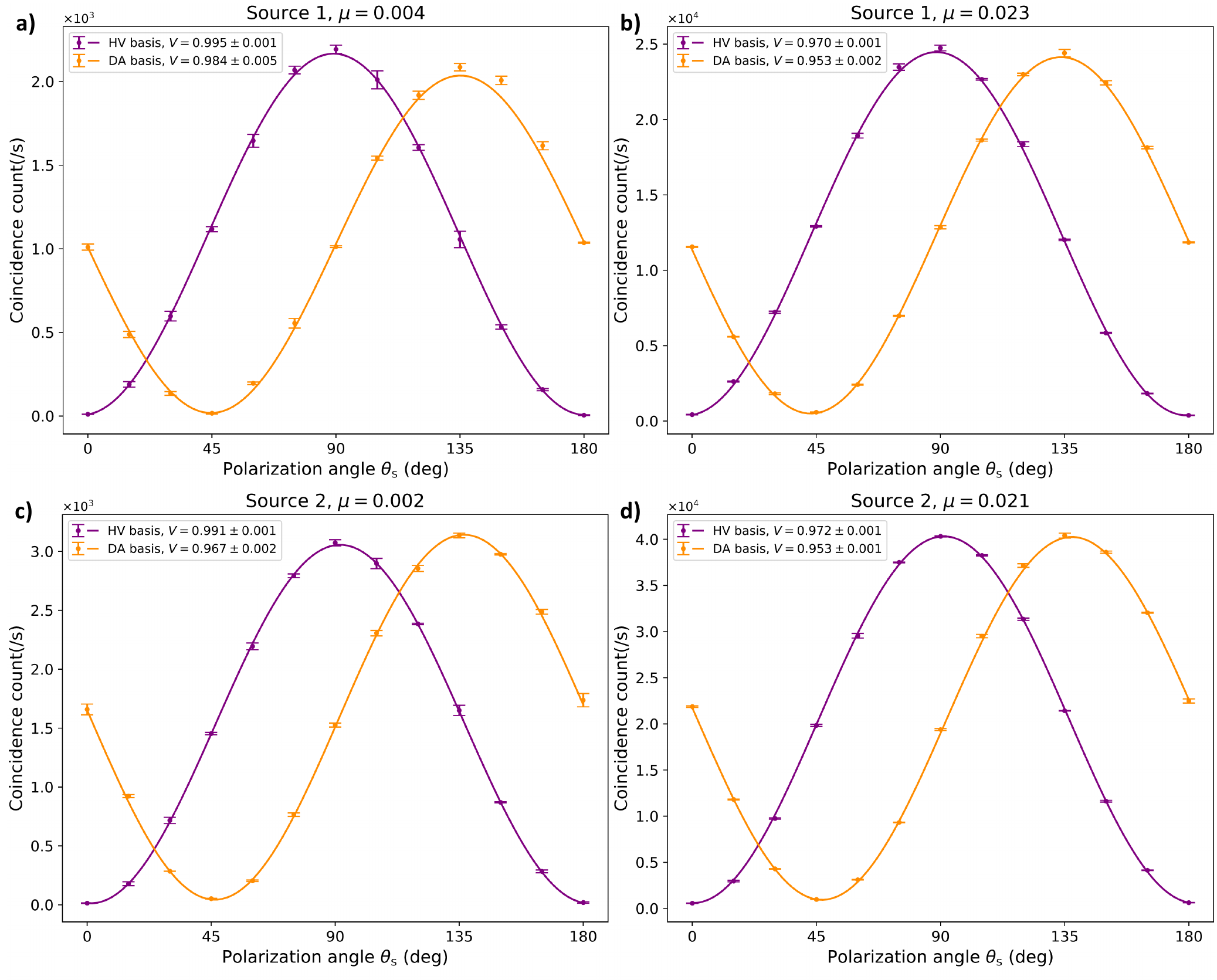}
\caption{
Experimental results of polarization-entanglement visibility with different average numbers of photon pairs.
The interference fringes are represented in purple for the rectilinear basis and in orange for the diagonal basis.
Subfigure (a) and (b) present the interference fringes of source 1 with $\mu = 0.004$ and $\mu = 0.023$, respectively.
Subfigure (c) and (d) present the entanglement visibility of source 2 with $\mu = 0.002$ and $\mu = 0.021$, respectively.
The polarization angle is twice the orientation angle of the fast axis of HWPs.
}
\label{visibility}
\end{figure*}

\begin{figure*}[t]
\centering
\includegraphics[width=.9\linewidth]{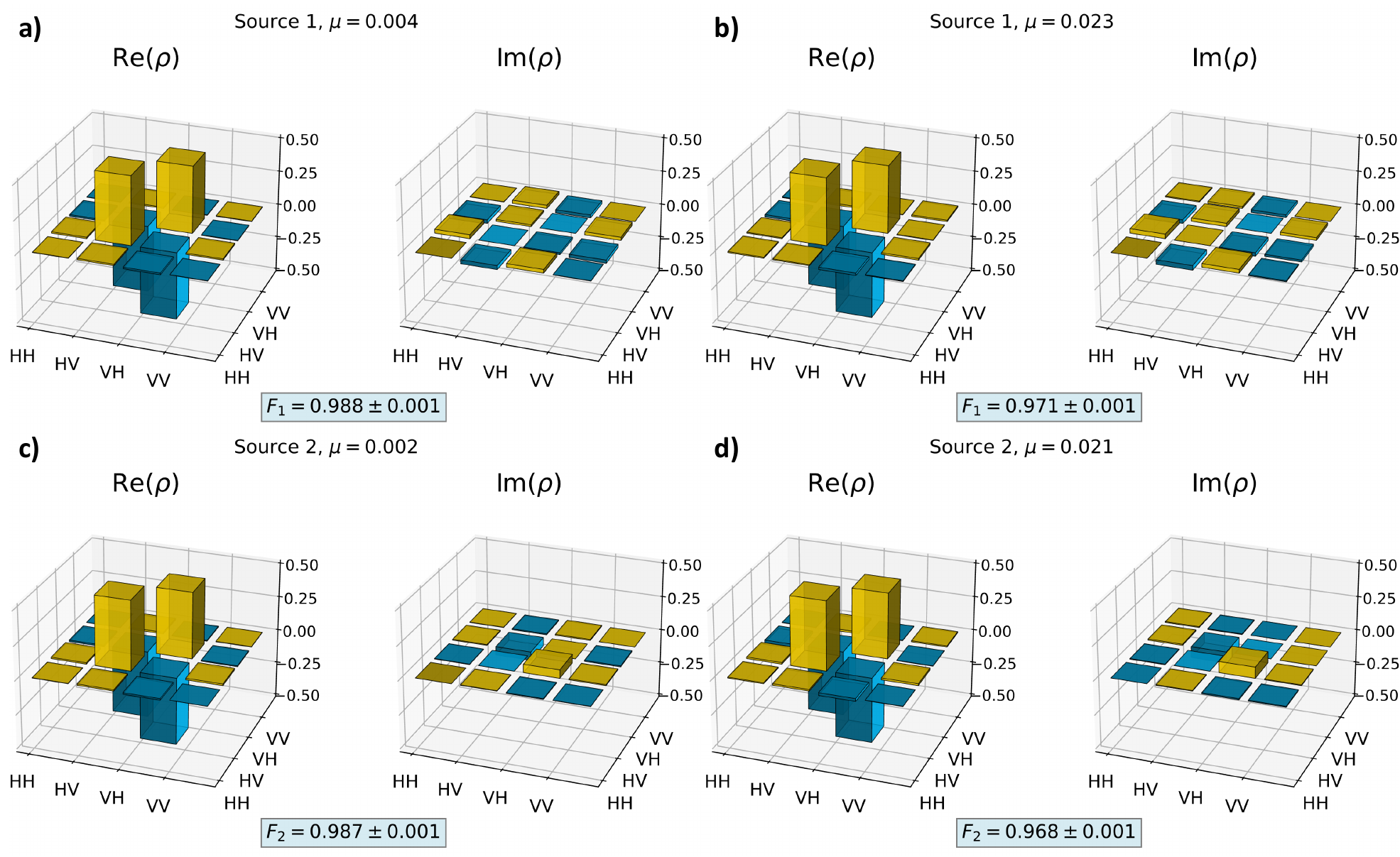}
\caption{
Quantum-state-tomography results of two entanglement sources under the different average numbers of photon pairs generated per pulse.
The matrix elements of the density matrices are represented in yellow for positive values and in blue for negative values.
Subfigures (a) and (b) present the density matrices of the entangled photon pairs generated by source 1 with $\mu=0.004$ and $\mu=0.023$, respectively.
Subfigures (c) and (d) present the quantum-state-tomography results of source 2 with $\mu=0.002$ and $\mu=0.021$, respectively.
$F$ represents the fidelity of the entangled photon pairs with respect to Bell state $\ket{\psi^{-}}$.
} 
\label{tomography}
\end{figure*}

\section{Performance of entanglement sources and SNSPD}
\label{performance}
The polarization-entanglement visibility and the fidelity can both evaluate the performance of the entanglement source.
Here, we provide the detailed experimental results of our entanglement sources with different average numbers of photon pairs generated per pulse.
The interference fringes of two entanglement sources with different average numbers of photon pairs are shown in Fig.~\ref{visibility}.
The polarization-entanglement visibility $V = (N_{\mathrm{max}} - N_{\mathrm{min}}) / (N_{\mathrm{max}} + N_{\mathrm{min}})$ can be calculated from the interference fringes, where $N_{\mathrm{max}}$ and $N_{\mathrm{min}}$ represent the maximum and minimum coincidence counts in the fringes, respectively.
The visibility of two sources could achieve 0.990 under the rectilinear basis and 0.965 under the diagonal basis with low average numbers of photon pairs. 
% ($\mu_1 = 0.004$, $\mu_2 = 0.002$).
With higher average numbers of photon pairs, which are used in the experimental demonstration of the source-independent QSS protocol, the visibility of two sources still exceeds 0.969 under the rectilinear basis and 0.951 under the diagonal basis.

For the fidelity of the photon pairs, we utilize quantum state tomography~\cite{james2001measurement} to construct the density matrix $\rho$ and calculate the fidelity $F=\left(\mathrm{tr} \sqrt{\sqrt{\rho} \sigma \sqrt{\rho}} \right)^2$ with respect to Bell state $\ket{\psi^{-}}$, where $\sigma = |\psi^{-}\rangle \langle\psi^{-}|$.
Fig.~\ref{tomography} presents the quantum-state-tomography results of polarization-entangled photon pairs generated by two entanglement sources. 
For each entanglement source under different $\mu$, we performed quantum state tomography three times and calculated the corresponding fidelity, thereby obtaining the mean and deviation of the fidelity.
In the experimental demonstration of the source-independent QSS protocol, the fidelity reaches 0.970 for source 1 with $\mu = 0.023$ and 0.967 for source 2 with $\mu = 0.021$, which presents the high quality of photon pairs.

In our experimental demonstration, 8 detection channels of SNSPD are required for the polarization measurement modules of the dealer and one player.
The detection efficiencies and dark count rates of 8 channels in our experimental setup are shown in Table~\ref{SNSPD_data}.
The average detection efficiency and average dark count rate are $0.83$ and $1.3 \times 10^{-7}$ per pulse, respectively, which are used as simulation parameters in the main text.

\begin{table}[t]
\renewcommand{\arraystretch}{1.2}
\centering
\caption{Main parameters of 8 detection channels of SNSPD. 
$\eta_d$ and $p_d$ denote the detection efficiency and dark count rate (per pulse), respectively. 
}
\begin{tabular*}{0.5\textwidth}{@{\extracolsep{\fill}}cccc}
\hline
Channel & $\eta_d$ & dark count(/s) & $p_d$(per pulse) \\
\hline
CH1 & 0.86 & 3 & $3.10\times10^{-8}$ \\
CH2 & 0.75 & 14 & $1.45\times10^{-7}$ \\
CH3 & 0.82 & 8 & $8.27\times10^{-8}$ \\
CH4 & 0.85 & 10 & $1.03\times10^{-7}$ \\
CH5 & 0.85 & 15 & $1.55\times10^{-7}$ \\
CH6 & 0.80 & 19 & $1.96\times10^{-7}$ \\
CH7 & 0.87 & 16 & $1.65\times10^{-7}$ \\
CH8 & 0.87 & 17 & $1.76\times10^{-7}$ \\
\hline
\end{tabular*}
\label{SNSPD_data}
\end{table}

\section{Experimental data for scalability and robustness}
\label{scalability and robustness}
\begin{figure*}[htbp]
\centering
\includegraphics[width=.8\linewidth]{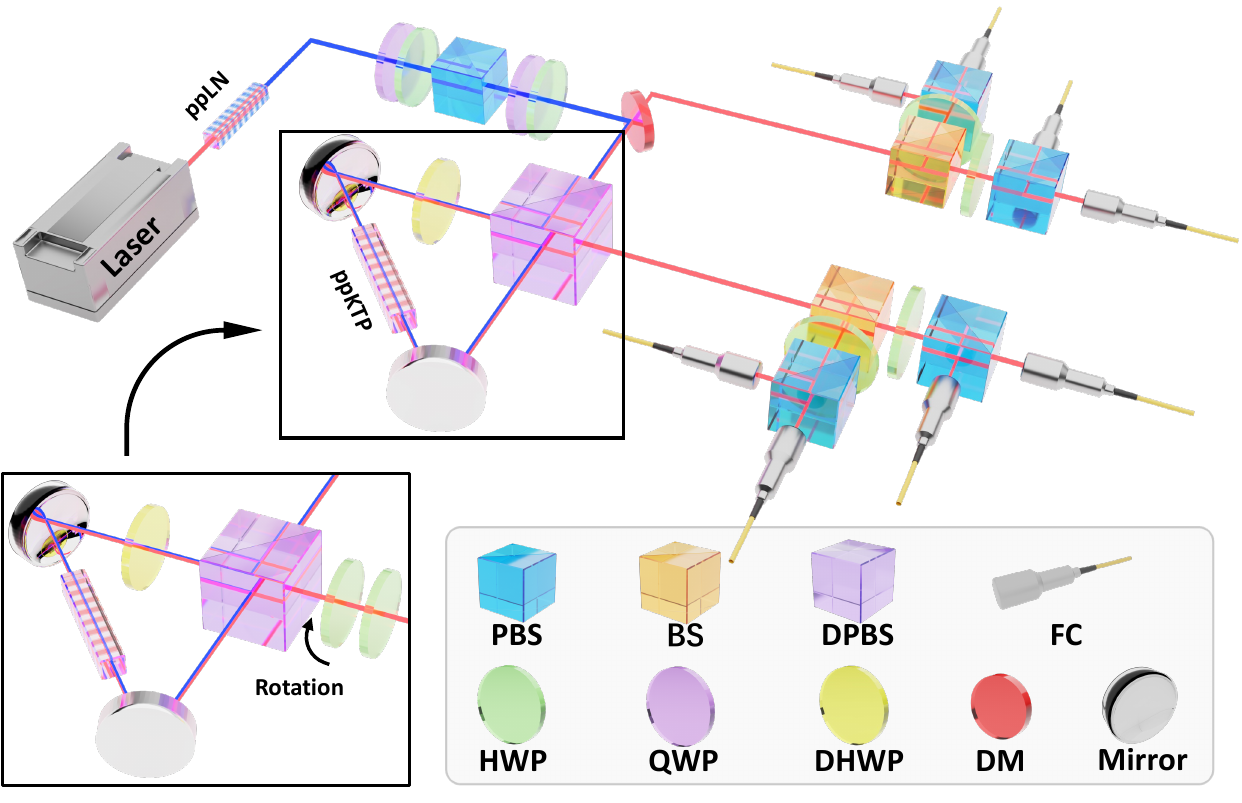}
\caption{Experimental setup for scalability and robustness.
For scalability, we utilize a single entanglement source and measurement results in different datasets to simulate the scenario with different numbers of users.
For robustness, we modify the entangled photon pairs by introducing two HWPs into the optical path, thereby changing the error rate.
PBS: polarization beam splitter; 
BS: beam splitter; 
DPBS: dichroic polarization beam splitter; 
QWP: quarter-wave plate; 
HWP: half-wave plate; 
DHWP: dichroic half-wave plate;
DM: dichroic mirror;
FC: fiber coulper.
} 
\label{setup_SM}
\end{figure*}

In this section, we present the experimental data demonstrating scalability with varying numbers of users and robustness under different error rates.
The experimental setup is shown in Fig.~\ref{setup_SM}.

As stated in the main text, we demonstrate the scalability of the resource-efficient SI QSS protocol using a single entanglement source and measurement results in different datasets.
Specifically, we utilize a single entanglement source to distribute entangled photon pairs and perform measurements under balanced basis selection, obtaining 9 datasets of equal duration.
Different datasets correspond to the measurement results between the dealer $A$ and different players $B_i$.
In each dataset, the number of pump pulses is $10^{11}$, the channel loss from the source to each user is 7.6 dB.
Then all measurement results are divided into three groups based on datasets, with each group containing 2, 3, and 4 datasets, corresponding to scenarios with 3, 4, and 5 users, respectively.
In each group, the measurement results in different datasets can be correlated to generate secure QSS keys through the postmatching method.
Table~\ref{experimental_results_3} shows the detailed error rates and QSS key rates involving 3, 4, and 5 users, respectively.
Note that since all the data originate from a single entanglement source, the error rates between dealer $A$ and each player $B_i$ are roughly the same.
The experimental QSS key rate decreases slightly as the number of participants increases, which demonstrates scalability of the resource-efficient SI QSS protocol.

To demonstrate the robustness under different error rates and present the source-independent performance, we modify the states generated by the entanglement source to change the error rate of our experimental system.
Similar to the experimental setup for demonstrating scalability, we use the data from a single entanglement source to simulate a network involving 3 users. 
By introducing two HWPs into the optical path, we rotate the polarization of one photon generated by the entanglement source, thereby changing the error rate and simulating a scenario where an eavesdropper manipulates the entanglement source.
Table~\ref{experimental_results_4} presents the experimental QSS key rates under different error rates.
The results indicate that as long as the error rate remains below the threshold, the resource-efficient SI QSS protocol can generate secure QSS keys, even if the entanglement source is manipulated by an eavesdropper. The greater the adversary’s control over the light source---or equivalently, the more information she acquires---the lower the achievable secure key rate. Crucially, however, any successfully distilled key maintains unconditional security. Notably, this security is entirely independent of any assumptions about the light source’s characteristics, relying solely on the measurement assumptions and outcomes.

\begin{table*}[htbp]
\renewcommand{\arraystretch}{1.5}
\centering
\caption{Summary of experimental data for the source-independent QSS protocol with different numbers of participants.
The key rates are tested with the same probabilities of basis selection ($p_x = 0.5$) and channel loss (7.6 dB).}
\begin{tabular*}{\textwidth}{@{\extracolsep{\fill}}ccccccccccc}
\hline
$n$ & $E_{AB_1}^X$ & $E_{AB_2}^X$ & $E_{AB_3}^X$ & $E_{AB_4}^X$ & $E^X$ & $E_{AB_1}^Z$ & $E_{AB_2}^Z$ & $E_{AB_3}^Z$ & $E_{AB_4}^Z$ & $r_{\mathrm{QSS}}$ (kbps) \\ 
\hline
3 & 0.0104 & 0.0102 & - & - & 0.0204 & 0.0265 & 0.0261 & - & - & $6.972$ \\
4 & 0.0104 & 0.0103 & 0.0101 & - & 0.0301 & 0.0262 & 0.0262 & 0.0261 & - & $6.461$ \\
5 & 0.0102 & 0.0102 & 0.0101 & 0.0103 & 0.0396 & 0.0263 & 0.0262 & 0.0261 & 0.0263 & $5.880$ \\
\hline
\end{tabular*}
\label{experimental_results_3}
\end{table*}

\begin{table*}[htbp]
\renewcommand{\arraystretch}{1.5}
\centering
\caption{Summary of experimental data for the robustness under different error rates.
The key rates are tested in three-user scenario with balanced basis selection ($p_x = 0.5$).}
\begin{tabular*}{\textwidth}{@{\extracolsep{\fill}}ccccccc}
\hline
$E_{AB_1}^X$ & $E_{AB_2}^X$ & $E^X$ & $E_{AB_1}^Z$ & $E_{AB_2}^Z$ & $\max_{j} \bar{\phi}^X_{AB_j}$ & $r_{\mathrm{QSS}}$ (kbps) \\ 
\hline
0.0104 & 0.0102 & 0.0204 & 0.0265 & 0.0261 & 0.0268 & $6.972$ \\
0.0212 & 0.0213 & 0.0416 & 0.0330 & 0.0331 & 0.0335 & $4.548$ \\
0.0318 & 0.0317 & 0.0615 & 0.0437 & 0.0431 & 0.0442 & $3.169$ \\
\hline
\end{tabular*}
\label{experimental_results_4}
\end{table*}

\clearpage
% \bibliography{Exp_SIQSS.bib}

\begin{thebibliography}{65}%
\makeatletter
\providecommand \@ifxundefined [1]{%
 \@ifx{#1\undefined}
}%
\providecommand \@ifnum [1]{%
 \ifnum #1\expandafter \@firstoftwo
 \else \expandafter \@secondoftwo
 \fi
}%
\providecommand \@ifx [1]{%
 \ifx #1\expandafter \@firstoftwo
 \else \expandafter \@secondoftwo
 \fi
}%
\providecommand \natexlab [1]{#1}%
\providecommand \enquote  [1]{``#1''}%
\providecommand \bibnamefont  [1]{#1}%
\providecommand \bibfnamefont [1]{#1}%
\providecommand \citenamefont [1]{#1}%
\providecommand \href@noop [0]{\@secondoftwo}%
\providecommand \href [0]{\begingroup \@sanitize@url \@href}%
\providecommand \@href[1]{\@@startlink{#1}\@@href}%
\providecommand \@@href[1]{\endgroup#1\@@endlink}%
\providecommand \@sanitize@url [0]{\catcode `\\12\catcode `\$12\catcode
  `\&12\catcode `\#12\catcode `\^12\catcode `\_12\catcode `\%12\relax}%
\providecommand \@@startlink[1]{}%
\providecommand \@@endlink[0]{}%
\providecommand \url  [0]{\begingroup\@sanitize@url \@url }%
\providecommand \@url [1]{\endgroup\@href {#1}{\urlprefix }}%
\providecommand \urlprefix  [0]{URL }%
\providecommand \Eprint [0]{\href }%
\providecommand \doibase [0]{https://doi.org/}%
\providecommand \selectlanguage [0]{\@gobble}%
\providecommand \bibinfo  [0]{\@secondoftwo}%
\providecommand \bibfield  [0]{\@secondoftwo}%
\providecommand \translation [1]{[#1]}%
\providecommand \BibitemOpen [0]{}%
\providecommand \bibitemStop [0]{}%
\providecommand \bibitemNoStop [0]{.\EOS\space}%
\providecommand \EOS [0]{\spacefactor3000\relax}%
\providecommand \BibitemShut  [1]{\csname bibitem#1\endcsname}%
\let\auto@bib@innerbib\@empty
%</preamble>
\bibitem [{\citenamefont {Xu}\ \emph {et~al.}(2020)\citenamefont {Xu},
  \citenamefont {Ma}, \citenamefont {Zhang}, \citenamefont {Lo},\ and\
  \citenamefont {Pan}}]{xu2020secure}%
  \BibitemOpen
  \bibfield  {author} {\bibinfo {author} {\bibfnamefont {F.}~\bibnamefont
  {Xu}}, \bibinfo {author} {\bibfnamefont {X.}~\bibnamefont {Ma}}, \bibinfo
  {author} {\bibfnamefont {Q.}~\bibnamefont {Zhang}}, \bibinfo {author}
  {\bibfnamefont {H.-K.}\ \bibnamefont {Lo}},\ and\ \bibinfo {author}
  {\bibfnamefont {J.-W.}\ \bibnamefont {Pan}},\ }\bibfield  {title} {\bibinfo
  {title} {Secure quantum key distribution with realistic devices},\
  }\href@noop {} {\bibfield  {journal} {\bibinfo  {journal} {Rev. Mod. Phys.}\
  }\textbf {\bibinfo {volume} {92}},\ \bibinfo {pages} {025002} (\bibinfo
  {year} {2020})}\BibitemShut {NoStop}%
\bibitem [{\citenamefont {Yin}\ \emph {et~al.}(2023)\citenamefont {Yin},
  \citenamefont {Fu}, \citenamefont {Li}, \citenamefont {Weng}, \citenamefont
  {Li}, \citenamefont {Gu}, \citenamefont {Lu}, \citenamefont {Huang},\ and\
  \citenamefont {Chen}}]{yin2023experimental}%
  \BibitemOpen
  \bibfield  {author} {\bibinfo {author} {\bibfnamefont {H.-L.}\ \bibnamefont
  {Yin}}, \bibinfo {author} {\bibfnamefont {Y.}~\bibnamefont {Fu}}, \bibinfo
  {author} {\bibfnamefont {C.-L.}\ \bibnamefont {Li}}, \bibinfo {author}
  {\bibfnamefont {C.-X.}\ \bibnamefont {Weng}}, \bibinfo {author}
  {\bibfnamefont {B.-H.}\ \bibnamefont {Li}}, \bibinfo {author} {\bibfnamefont
  {J.}~\bibnamefont {Gu}}, \bibinfo {author} {\bibfnamefont {Y.-S.}\
  \bibnamefont {Lu}}, \bibinfo {author} {\bibfnamefont {S.}~\bibnamefont
  {Huang}},\ and\ \bibinfo {author} {\bibfnamefont {Z.-B.}\ \bibnamefont
  {Chen}},\ }\bibfield  {title} {\bibinfo {title} {Experimental quantum secure
  network with digital signatures and encryption},\ }\href@noop {} {\bibfield
  {journal} {\bibinfo  {journal} {Natl. Sci. Rev.}\ }\textbf {\bibinfo {volume}
  {10}},\ \bibinfo {pages} {nwac228} (\bibinfo {year} {2023})}\BibitemShut
  {NoStop}%
\bibitem [{\citenamefont {Azuma}\ \emph {et~al.}(2023)\citenamefont {Azuma},
  \citenamefont {Economou}, \citenamefont {Elkouss}, \citenamefont {Hilaire},
  \citenamefont {Jiang}, \citenamefont {Lo},\ and\ \citenamefont
  {Tzitrin}}]{azuma2023quantum}%
  \BibitemOpen
  \bibfield  {author} {\bibinfo {author} {\bibfnamefont {K.}~\bibnamefont
  {Azuma}}, \bibinfo {author} {\bibfnamefont {S.~E.}\ \bibnamefont {Economou}},
  \bibinfo {author} {\bibfnamefont {D.}~\bibnamefont {Elkouss}}, \bibinfo
  {author} {\bibfnamefont {P.}~\bibnamefont {Hilaire}}, \bibinfo {author}
  {\bibfnamefont {L.}~\bibnamefont {Jiang}}, \bibinfo {author} {\bibfnamefont
  {H.-K.}\ \bibnamefont {Lo}},\ and\ \bibinfo {author} {\bibfnamefont
  {I.}~\bibnamefont {Tzitrin}},\ }\bibfield  {title} {\bibinfo {title} {Quantum
  repeaters: From quantum networks to the quantum internet},\ }\href@noop {}
  {\bibfield  {journal} {\bibinfo  {journal} {Rev. Mod. Phys.}\ }\textbf
  {\bibinfo {volume} {95}},\ \bibinfo {pages} {045006} (\bibinfo {year}
  {2023})}\BibitemShut {NoStop}%
\bibitem [{\citenamefont {Li}\ \emph {et~al.}(2024)\citenamefont {Li},
  \citenamefont {Yin},\ and\ \citenamefont {Chen}}]{li2024asynchronous}%
  \BibitemOpen
  \bibfield  {author} {\bibinfo {author} {\bibfnamefont {C.-L.}\ \bibnamefont
  {Li}}, \bibinfo {author} {\bibfnamefont {H.-L.}\ \bibnamefont {Yin}},\ and\
  \bibinfo {author} {\bibfnamefont {Z.-B.}\ \bibnamefont {Chen}},\ }\bibfield
  {title} {\bibinfo {title} {Asynchronous quantum repeater using multiple
  quantum memory},\ }\href@noop {} {\bibfield  {journal} {\bibinfo  {journal}
  {Rep. Prog. Phys.}\ }\textbf {\bibinfo {volume} {87}},\ \bibinfo {pages}
  {127901} (\bibinfo {year} {2024})}\BibitemShut {NoStop}%
\bibitem [{\citenamefont {Zhou}\ \emph {et~al.}(2023)\citenamefont {Zhou},
  \citenamefont {Lin}, \citenamefont {Xie}, \citenamefont {Lu}, \citenamefont
  {Jing}, \citenamefont {Yin},\ and\ \citenamefont
  {Yuan}}]{Zhou2023OvercomeRateloss}%
  \BibitemOpen
  \bibfield  {author} {\bibinfo {author} {\bibfnamefont {L.}~\bibnamefont
  {Zhou}}, \bibinfo {author} {\bibfnamefont {J.}~\bibnamefont {Lin}}, \bibinfo
  {author} {\bibfnamefont {Y.-M.}\ \bibnamefont {Xie}}, \bibinfo {author}
  {\bibfnamefont {Y.-S.}\ \bibnamefont {Lu}}, \bibinfo {author} {\bibfnamefont
  {Y.}~\bibnamefont {Jing}}, \bibinfo {author} {\bibfnamefont {H.-L.}\
  \bibnamefont {Yin}},\ and\ \bibinfo {author} {\bibfnamefont {Z.}~\bibnamefont
  {Yuan}},\ }\bibfield  {title} {\bibinfo {title} {Experimental quantum
  communication overcomes the rate-loss limit without global phase tracking},\
  }\href@noop {} {\bibfield  {journal} {\bibinfo  {journal} {Phys. Rev. Lett.}\
  }\textbf {\bibinfo {volume} {130}},\ \bibinfo {pages} {250801} (\bibinfo
  {year} {2023})}\BibitemShut {NoStop}%
\bibitem [{\citenamefont {Xie}\ \emph {et~al.}(2022)\citenamefont {Xie},
  \citenamefont {Lu}, \citenamefont {Weng}, \citenamefont {Cao}, \citenamefont
  {Jia}, \citenamefont {Bao}, \citenamefont {Wang}, \citenamefont {Fu},
  \citenamefont {Yin},\ and\ \citenamefont {Chen}}]{xie2022breaking}%
  \BibitemOpen
  \bibfield  {author} {\bibinfo {author} {\bibfnamefont {Y.-M.}\ \bibnamefont
  {Xie}}, \bibinfo {author} {\bibfnamefont {Y.-S.}\ \bibnamefont {Lu}},
  \bibinfo {author} {\bibfnamefont {C.-X.}\ \bibnamefont {Weng}}, \bibinfo
  {author} {\bibfnamefont {X.-Y.}\ \bibnamefont {Cao}}, \bibinfo {author}
  {\bibfnamefont {Z.-Y.}\ \bibnamefont {Jia}}, \bibinfo {author} {\bibfnamefont
  {Y.}~\bibnamefont {Bao}}, \bibinfo {author} {\bibfnamefont {Y.}~\bibnamefont
  {Wang}}, \bibinfo {author} {\bibfnamefont {Y.}~\bibnamefont {Fu}}, \bibinfo
  {author} {\bibfnamefont {H.-L.}\ \bibnamefont {Yin}},\ and\ \bibinfo {author}
  {\bibfnamefont {Z.-B.}\ \bibnamefont {Chen}},\ }\bibfield  {title} {\bibinfo
  {title} {Breaking the rate-loss bound of quantum key distribution with
  asynchronous two-photon interference},\ }\href@noop {} {\bibfield  {journal}
  {\bibinfo  {journal} {PRX Quantum}\ }\textbf {\bibinfo {volume} {3}},\
  \bibinfo {pages} {020315} (\bibinfo {year} {2022})}\BibitemShut {NoStop}%
\bibitem [{\citenamefont {Zeng}\ \emph {et~al.}(2022)\citenamefont {Zeng},
  \citenamefont {Zhou}, \citenamefont {Wu},\ and\ \citenamefont
  {Ma}}]{zeng2022mode}%
  \BibitemOpen
  \bibfield  {author} {\bibinfo {author} {\bibfnamefont {P.}~\bibnamefont
  {Zeng}}, \bibinfo {author} {\bibfnamefont {H.}~\bibnamefont {Zhou}}, \bibinfo
  {author} {\bibfnamefont {W.}~\bibnamefont {Wu}},\ and\ \bibinfo {author}
  {\bibfnamefont {X.}~\bibnamefont {Ma}},\ }\bibfield  {title} {\bibinfo
  {title} {Mode-pairing quantum key distribution},\ }\href@noop {} {\bibfield
  {journal} {\bibinfo  {journal} {Nat. Commun.}\ }\textbf {\bibinfo {volume}
  {13}},\ \bibinfo {pages} {3903} (\bibinfo {year} {2022})}\BibitemShut
  {NoStop}%
\bibitem [{\citenamefont {Kimble}(2008)}]{kimble2008quantum}%
  \BibitemOpen
  \bibfield  {author} {\bibinfo {author} {\bibfnamefont {H.~J.}\ \bibnamefont
  {Kimble}},\ }\bibfield  {title} {\bibinfo {title} {The quantum internet},\
  }\href@noop {} {\bibfield  {journal} {\bibinfo  {journal} {Nature (London)}\ }\textbf
  {\bibinfo {volume} {453}},\ \bibinfo {pages} {1023} (\bibinfo {year}
  {2008})}\BibitemShut {NoStop}%
\bibitem [{\citenamefont {Wehner}\ \emph {et~al.}(2018)\citenamefont {Wehner},
  \citenamefont {Elkouss},\ and\ \citenamefont {Hanson}}]{wehner2018quantum}%
  \BibitemOpen
  \bibfield  {author} {\bibinfo {author} {\bibfnamefont {S.}~\bibnamefont
  {Wehner}}, \bibinfo {author} {\bibfnamefont {D.}~\bibnamefont {Elkouss}},\
  and\ \bibinfo {author} {\bibfnamefont {R.}~\bibnamefont {Hanson}},\
  }\bibfield  {title} {\bibinfo {title} {Quantum internet: A vision for the
  road ahead},\ }\href@noop {} {\bibfield  {journal} {\bibinfo  {journal}
  {Science}\ }\textbf {\bibinfo {volume} {362}},\ \bibinfo {pages} {eaam9288}
  (\bibinfo {year} {2018})}\BibitemShut {NoStop}%
\bibitem [{\citenamefont {Shamir}(1979)}]{Shamir1979SecretSharing}%
  \BibitemOpen
  \bibfield  {author} {\bibinfo {author} {\bibfnamefont {A.}~\bibnamefont
  {Shamir}},\ }\bibfield  {title} {\bibinfo {title} {How to share a secret},\
  }\href@noop {} {\bibfield  {journal} {\bibinfo  {journal} {Commun. ACM}\
  }\textbf {\bibinfo {volume} {22}},\ \bibinfo {pages} {612} (\bibinfo {year}
  {1979})}\BibitemShut {NoStop}%
\bibitem [{\citenamefont {Blakley}(1979)}]{Blakley1979SecretSharing}%
  \BibitemOpen
  \bibfield  {author} {\bibinfo {author} {\bibfnamefont {G.~R.}\ \bibnamefont
  {Blakley}},\ }\bibfield  {title} {\bibinfo {title} {Safeguarding
  cryptographic keys},\ }in\ \href@noop {} {\emph {\bibinfo {booktitle}
  {Managing Requirements Knowledge, International Workshop on}}}\ (\bibinfo
  {organization} {IEEE Computer Society},\ \bibinfo {year} {1979})\ p.\
  \bibinfo {pages} {313}\BibitemShut {NoStop}%
\bibitem [{\citenamefont {Hillery}\ \emph {et~al.}(1999)\citenamefont
  {Hillery}, \citenamefont {Bu{\v{z}}ek},\ and\ \citenamefont
  {Berthiaume}}]{Hillery1999QSS}%
  \BibitemOpen
  \bibfield  {author} {\bibinfo {author} {\bibfnamefont {M.}~\bibnamefont
  {Hillery}}, \bibinfo {author} {\bibfnamefont {V.}~\bibnamefont
  {Bu{\v{z}}ek}},\ and\ \bibinfo {author} {\bibfnamefont {A.}~\bibnamefont
  {Berthiaume}},\ }\bibfield  {title} {\bibinfo {title} {Quantum secret
  sharing},\ }\href@noop {} {\bibfield  {journal} {\bibinfo  {journal} {Phys.
  Rev. A}\ }\textbf {\bibinfo {volume} {59}},\ \bibinfo {pages} {1829}
  (\bibinfo {year} {1999})}\BibitemShut {NoStop}%
\bibitem [{\citenamefont {Cleve}\ \emph {et~al.}(1999)\citenamefont {Cleve},
  \citenamefont {Gottesman},\ and\ \citenamefont {Lo}}]{Cleve1999QSS}%
  \BibitemOpen
  \bibfield  {author} {\bibinfo {author} {\bibfnamefont {R.}~\bibnamefont
  {Cleve}}, \bibinfo {author} {\bibfnamefont {D.}~\bibnamefont {Gottesman}},\
  and\ \bibinfo {author} {\bibfnamefont {H.-K.}\ \bibnamefont {Lo}},\
  }\bibfield  {title} {\bibinfo {title} {How to share a quantum secret},\
  }\href@noop {} {\bibfield  {journal} {\bibinfo  {journal} {Phys. Rev. Lett.}\
  }\textbf {\bibinfo {volume} {83}},\ \bibinfo {pages} {648} (\bibinfo {year}
  {1999})}\BibitemShut {NoStop}%
\bibitem [{\citenamefont {Fu}\ \emph {et~al.}(2015)\citenamefont {Fu},
  \citenamefont {Yin}, \citenamefont {Chen},\ and\ \citenamefont
  {Chen}}]{Fu2015MDIQSSQCKA}%
  \BibitemOpen
  \bibfield  {author} {\bibinfo {author} {\bibfnamefont {Y.}~\bibnamefont
  {Fu}}, \bibinfo {author} {\bibfnamefont {H.-L.}\ \bibnamefont {Yin}},
  \bibinfo {author} {\bibfnamefont {T.-Y.}\ \bibnamefont {Chen}},\ and\
  \bibinfo {author} {\bibfnamefont {Z.-B.}\ \bibnamefont {Chen}},\ }\bibfield
  {title} {\bibinfo {title} {Long-distance measurement-device-independent
  multiparty quantum communication},\ }\href@noop {} {\bibfield  {journal}
  {\bibinfo  {journal} {Phys. Rev. Lett.}\ }\textbf {\bibinfo {volume} {114}},\
  \bibinfo {pages} {090501} (\bibinfo {year} {2015})}\BibitemShut {NoStop}%
\bibitem [{\citenamefont {Gu}\ \emph {et~al.}(2021)\citenamefont {Gu},
  \citenamefont {Xie}, \citenamefont {Liu}, \citenamefont {Fu}, \citenamefont
  {Yin},\ and\ \citenamefont {Chen}}]{gu2021secure}%
  \BibitemOpen
  \bibfield  {author} {\bibinfo {author} {\bibfnamefont {J.}~\bibnamefont
  {Gu}}, \bibinfo {author} {\bibfnamefont {Y.-M.}\ \bibnamefont {Xie}},
  \bibinfo {author} {\bibfnamefont {W.-B.}\ \bibnamefont {Liu}}, \bibinfo
  {author} {\bibfnamefont {Y.}~\bibnamefont {Fu}}, \bibinfo {author}
  {\bibfnamefont {H.-L.}\ \bibnamefont {Yin}},\ and\ \bibinfo {author}
  {\bibfnamefont {Z.-B.}\ \bibnamefont {Chen}},\ }\bibfield  {title} {\bibinfo
  {title} {Secure quantum secret sharing without signal disturbance
  monitoring},\ }\href@noop {} {\bibfield  {journal} {\bibinfo  {journal} {Opt.
  Express}\ }\textbf {\bibinfo {volume} {29}},\ \bibinfo {pages} {32244}
  (\bibinfo {year} {2021})}\BibitemShut {NoStop}%
\bibitem [{\citenamefont {Wang}\ \emph {et~al.}(2024)\citenamefont {Wang},
  \citenamefont {Sun}, \citenamefont {Cao}, \citenamefont {Yin},\ and\
  \citenamefont {Chen}}]{wang2024experimental}%
  \BibitemOpen
  \bibfield  {author} {\bibinfo {author} {\bibfnamefont {Y.-Z.}\ \bibnamefont
  {Wang}}, \bibinfo {author} {\bibfnamefont {X.-R.}\ \bibnamefont {Sun}},
  \bibinfo {author} {\bibfnamefont {X.-Y.}\ \bibnamefont {Cao}}, \bibinfo
  {author} {\bibfnamefont {H.-L.}\ \bibnamefont {Yin}},\ and\ \bibinfo {author}
  {\bibfnamefont {Z.-B.}\ \bibnamefont {Chen}},\ }\bibfield  {title} {\bibinfo
  {title} {Experimental coherent-state quantum secret sharing with finite
  pulses},\ }\href@noop {} {\bibfield  {journal} {\bibinfo  {journal} {Phys.
  Rev. Appl.}\ }\textbf {\bibinfo {volume} {22}},\ \bibinfo {pages} {044018}
  (\bibinfo {year} {2024})}\BibitemShut {NoStop}%
\bibitem [{\citenamefont {Lance}\ \emph {et~al.}(2004)\citenamefont {Lance},
  \citenamefont {Symul}, \citenamefont {Bowen}, \citenamefont {Sanders},\ and\
  \citenamefont {Lam}}]{lance2004tripartite}%
  \BibitemOpen
  \bibfield  {author} {\bibinfo {author} {\bibfnamefont {A.~M.}\ \bibnamefont
  {Lance}}, \bibinfo {author} {\bibfnamefont {T.}~\bibnamefont {Symul}},
  \bibinfo {author} {\bibfnamefont {W.~P.}\ \bibnamefont {Bowen}}, \bibinfo
  {author} {\bibfnamefont {B.~C.}\ \bibnamefont {Sanders}},\ and\ \bibinfo
  {author} {\bibfnamefont {P.~K.}\ \bibnamefont {Lam}},\ }\bibfield  {title}
  {\bibinfo {title} {Tripartite quantum state sharing},\ }\href@noop {}
  {\bibfield  {journal} {\bibinfo  {journal} {Phys. Rev. Lett.}\ }\textbf
  {\bibinfo {volume} {92}},\ \bibinfo {pages} {177903} (\bibinfo {year}
  {2004})}\BibitemShut {NoStop}%
\bibitem [{\citenamefont {Singh}\ and\ \citenamefont
  {Srikanth}(2005)}]{Singh2005GeneralizedQSS}%
  \BibitemOpen
  \bibfield  {author} {\bibinfo {author} {\bibfnamefont {S.~Kumar}\ \bibnamefont
  {Singh}}\ and\ \bibinfo {author} {\bibfnamefont {R.}~\bibnamefont
  {Srikanth}},\ }\bibfield  {title} {\bibinfo {title} {Generalized quantum
  secret sharing},\ }\href@noop {} {\bibfield  {journal} {\bibinfo  {journal}
  {Phys. Rev. A}\ }\textbf {\bibinfo {volume} {71}},\ \bibinfo {pages} {012328}
  (\bibinfo {year} {2005})}\BibitemShut {NoStop}%
\bibitem [{\citenamefont {Yu}\ \emph {et~al.}(2008)\citenamefont {Yu},
  \citenamefont {Lin},\ and\ \citenamefont {Huang}}]{Yu2008MultilevelQSS}%
  \BibitemOpen
  \bibfield  {author} {\bibinfo {author} {\bibfnamefont {I.-Ching}\ \bibnamefont
  {Yu}}, \bibinfo {author} {\bibfnamefont {F.-L.}\ \bibnamefont {Lin}},\ and\
  \bibinfo {author} {\bibfnamefont {C.-Y.}\ \bibnamefont {Huang}},\ }\bibfield
  {title} {\bibinfo {title} {Quantum secret sharing with multilevel mutually
  (un)biased bases},\ }\href@noop {} {\bibfield  {journal} {\bibinfo  {journal}
  {Phys. Rev. A}\ }\textbf {\bibinfo {volume} {78}},\ \bibinfo {pages} {012344}
  (\bibinfo {year} {2008})}\BibitemShut {NoStop}%
\bibitem [{\citenamefont {Markham}\ and\ \citenamefont
  {Sanders}(2008)}]{Markham2008GraphQSS}%
  \BibitemOpen
  \bibfield  {author} {\bibinfo {author} {\bibfnamefont {D.}~\bibnamefont
  {Markham}}\ and\ \bibinfo {author} {\bibfnamefont {B.~C.}\ \bibnamefont
  {Sanders}},\ }\bibfield  {title} {\bibinfo {title} {Graph states for quantum
  secret sharing},\ }\href@noop {} {\bibfield  {journal} {\bibinfo  {journal}
  {Phys. Rev. A}\ }\textbf {\bibinfo {volume} {78}},\ \bibinfo {pages} {042309}
  (\bibinfo {year} {2008})}\BibitemShut {NoStop}%
\bibitem [{\citenamefont {Cai}\ \emph {et~al.}(2017)\citenamefont {Cai},
  \citenamefont {Roslund}, \citenamefont {Ferrini}, \citenamefont {Arzani},
  \citenamefont {Xu}, \citenamefont {Fabre},\ and\ \citenamefont
  {Treps}}]{Cai2017MultimodeGraphQSS}%
  \BibitemOpen
  \bibfield  {author} {\bibinfo {author} {\bibfnamefont {Y.}~\bibnamefont
  {Cai}}, \bibinfo {author} {\bibfnamefont {J.}~\bibnamefont {Roslund}},
  \bibinfo {author} {\bibfnamefont {G.}~\bibnamefont {Ferrini}}, \bibinfo
  {author} {\bibfnamefont {F.}~\bibnamefont {Arzani}}, \bibinfo {author}
  {\bibfnamefont {X.}~\bibnamefont {Xu}}, \bibinfo {author} {\bibfnamefont
  {C.}~\bibnamefont {Fabre}},\ and\ \bibinfo {author} {\bibfnamefont
  {N.}~\bibnamefont {Treps}},\ }\bibfield  {title} {\bibinfo {title} {Multimode
  entanglement in reconfigurable graph states using optical frequency combs},\
  }\href@noop {} {\bibfield  {journal} {\bibinfo  {journal} {Nat. Commun.}\
  }\textbf {\bibinfo {volume} {8}},\ \bibinfo {pages} {15645} (\bibinfo {year}
  {2017})}\BibitemShut {NoStop}%
\bibitem [{\citenamefont {De~Oliveira}\ \emph {et~al.}(2020)\citenamefont
  {De~Oliveira}, \citenamefont {Nape}, \citenamefont {Pinnell}, \citenamefont
  {TabeBordbar},\ and\ \citenamefont {Forbes}}]{de2020experimental}%
  \BibitemOpen
  \bibfield  {author} {\bibinfo {author} {\bibfnamefont {M.}~\bibnamefont
  {de~Oliveira}}, \bibinfo {author} {\bibfnamefont {I.}~\bibnamefont {Nape}},
  \bibinfo {author} {\bibfnamefont {J.}~\bibnamefont {Pinnell}}, \bibinfo
  {author} {\bibfnamefont {N.}~\bibnamefont {TabeBordbar}},\ and\ \bibinfo
  {author} {\bibfnamefont {A.}~\bibnamefont {Forbes}},\ }\bibfield  {title}
  {\bibinfo {title} {Experimental high-dimensional quantum secret sharing with
  spin-orbit-structured photons},\ }\href@noop {} {\bibfield  {journal}
  {\bibinfo  {journal} {Phys. Rev. A}\ }\textbf {\bibinfo {volume} {101}},\
  \bibinfo {pages} {042303} (\bibinfo {year} {2020})}\BibitemShut {NoStop}%
\bibitem [{\citenamefont {Tavakoli}\ \emph {et~al.}(2015)\citenamefont
  {Tavakoli}, \citenamefont {Herbauts}, \citenamefont {{\.Z}ukowski},\ and\
  \citenamefont {Bourennane}}]{Tavakoli2015DLevelQSS}%
  \BibitemOpen
  \bibfield  {author} {\bibinfo {author} {\bibfnamefont {A.}~\bibnamefont
  {Tavakoli}}, \bibinfo {author} {\bibfnamefont {I.}~\bibnamefont {Herbauts}},
  \bibinfo {author} {\bibfnamefont {M.}~\bibnamefont {{\.Z}ukowski}},\ and\
  \bibinfo {author} {\bibfnamefont {M.}~\bibnamefont {Bourennane}},\ }\bibfield
   {title} {\bibinfo {title} {Secret sharing with a single $d$-level quantum
  system},\ }\href@noop {} {\bibfield  {journal} {\bibinfo  {journal} {Phys.
  Rev. A}\ }\textbf {\bibinfo {volume} {92}},\ \bibinfo {pages} {030302(R)}
  (\bibinfo {year} {2015})}\BibitemShut {NoStop}%
\bibitem [{\citenamefont {Li}\ \emph {et~al.}(2023)\citenamefont {Li},
  \citenamefont {Fu}, \citenamefont {Liu}, \citenamefont {Xie}, \citenamefont
  {Li}, \citenamefont {Zhou}, \citenamefont {Yin},\ and\ \citenamefont
  {Chen}}]{Li2023AMDIQSS}%
  \BibitemOpen
  \bibfield  {author} {\bibinfo {author} {\bibfnamefont {C.-L.}\ \bibnamefont
  {Li}}, \bibinfo {author} {\bibfnamefont {Y.}~\bibnamefont {Fu}}, \bibinfo
  {author} {\bibfnamefont {W.-B.}\ \bibnamefont {Liu}}, \bibinfo {author}
  {\bibfnamefont {Y.-M.}\ \bibnamefont {Xie}}, \bibinfo {author} {\bibfnamefont
  {B.-H.}\ \bibnamefont {Li}}, \bibinfo {author} {\bibfnamefont {M.-G.}\
  \bibnamefont {Zhou}}, \bibinfo {author} {\bibfnamefont {H.-L.}\ \bibnamefont
  {Yin}},\ and\ \bibinfo {author} {\bibfnamefont {Z.-B.}\ \bibnamefont
  {Chen}},\ }\bibfield  {title} {\bibinfo {title} {Breaking the rate-distance
  limitation of measurement-device-independent quantum secret sharing},\
  }\href@noop {} {\bibfield  {journal} {\bibinfo  {journal} {Phys. Rev. Res.}\
  }\textbf {\bibinfo {volume} {5}},\ \bibinfo {pages} {033077} (\bibinfo {year}
  {2023})}\BibitemShut {NoStop}%
\bibitem [{\citenamefont {Zhang}\ \emph {et~al.}(2024)\citenamefont {Zhang},
  \citenamefont {Zhong}, \citenamefont {Du}, \citenamefont {Shen},
  \citenamefont {Li}, \citenamefont {Zhang}, \citenamefont {Zhou},\ and\
  \citenamefont {Sheng}}]{zhang2024device}%
  \BibitemOpen
  \bibfield  {author} {\bibinfo {author} {\bibfnamefont {Q.}~\bibnamefont
  {Zhang}}, \bibinfo {author} {\bibfnamefont {W.}~\bibnamefont {Zhong}},
  \bibinfo {author} {\bibfnamefont {M.-M.}\ \bibnamefont {Du}}, \bibinfo
  {author} {\bibfnamefont {S.-T.}\ \bibnamefont {Shen}}, \bibinfo {author}
  {\bibfnamefont {X.-Y.}\ \bibnamefont {Li}}, \bibinfo {author} {\bibfnamefont
  {A.-L.}\ \bibnamefont {Zhang}}, \bibinfo {author} {\bibfnamefont
  {L.}~\bibnamefont {Zhou}},\ and\ \bibinfo {author} {\bibfnamefont {Y.-B.}\
  \bibnamefont {Sheng}},\ }\bibfield  {title} {\bibinfo {title}
  {Device-independent quantum secret sharing with noise preprocessing and
  postselection},\ }\href@noop {} {\bibfield  {journal} {\bibinfo  {journal}
  {Phys. Rev. A}\ }\textbf {\bibinfo {volume} {110}},\ \bibinfo {pages}
  {042403} (\bibinfo {year} {2024})}\BibitemShut {NoStop}%
\bibitem [{\citenamefont {Schmid}\ \emph {et~al.}(2005)\citenamefont {Schmid},
  \citenamefont {Trojek}, \citenamefont {Bourennane}, \citenamefont
  {Kurtsiefer}, \citenamefont {{\.Z}ukowski},\ and\ \citenamefont
  {Weinfurter}}]{schmid2005experimental}%
  \BibitemOpen
  \bibfield  {author} {\bibinfo {author} {\bibfnamefont {C.}~\bibnamefont
  {Schmid}}, \bibinfo {author} {\bibfnamefont {P.}~\bibnamefont {Trojek}},
  \bibinfo {author} {\bibfnamefont {M.}~\bibnamefont {Bourennane}}, \bibinfo
  {author} {\bibfnamefont {C.}~\bibnamefont {Kurtsiefer}}, \bibinfo {author}
  {\bibfnamefont {M.}~\bibnamefont {{\.Z}ukowski}},\ and\ \bibinfo {author}
  {\bibfnamefont {H.}~\bibnamefont {Weinfurter}},\ }\bibfield  {title}
  {\bibinfo {title} {Experimental single qubit quantum secret sharing},\
  }\href@noop {} {\bibfield  {journal} {\bibinfo  {journal} {Phys. Rev. Lett.}\
  }\textbf {\bibinfo {volume} {95}},\ \bibinfo {pages} {230505} (\bibinfo
  {year} {2005})}\BibitemShut {NoStop}%
\bibitem [{\citenamefont {Chen}\ \emph {et~al.}(2005)\citenamefont {Chen},
  \citenamefont {Zhang}, \citenamefont {Zhao}, \citenamefont {Zhou},
  \citenamefont {Lu}, \citenamefont {Peng}, \citenamefont {Yang},\ and\
  \citenamefont {Pan}}]{chen2005experimental}%
  \BibitemOpen
  \bibfield  {author} {\bibinfo {author} {\bibfnamefont {Y.-A.}\ \bibnamefont
  {Chen}}, \bibinfo {author} {\bibfnamefont {A.-N.}\ \bibnamefont {Zhang}},
  \bibinfo {author} {\bibfnamefont {Z.}~\bibnamefont {Zhao}}, \bibinfo {author}
  {\bibfnamefont {X.-Q.}\ \bibnamefont {Zhou}}, \bibinfo {author}
  {\bibfnamefont {C.-Y.}\ \bibnamefont {Lu}}, \bibinfo {author} {\bibfnamefont
  {C.-Z.}\ \bibnamefont {Peng}}, \bibinfo {author} {\bibfnamefont
  {T.}~\bibnamefont {Yang}},\ and\ \bibinfo {author} {\bibfnamefont {J.-W.}\
  \bibnamefont {Pan}},\ }\bibfield  {title} {\bibinfo {title} {Experimental
  quantum secret sharing and third-man quantum cryptography},\ }\href@noop {}
  {\bibfield  {journal} {\bibinfo  {journal} {Phys. Rev. Lett.}\ }\textbf
  {\bibinfo {volume} {95}},\ \bibinfo {pages} {200502} (\bibinfo {year}
  {2005})}\BibitemShut {NoStop}%
\bibitem [{\citenamefont {Gaertner}\ \emph {et~al.}(2007)\citenamefont
  {Gaertner}, \citenamefont {Kurtsiefer}, \citenamefont {Bourennane},\ and\
  \citenamefont {Weinfurter}}]{Gaertner2007FourPartyQSS}%
  \BibitemOpen
  \bibfield  {author} {\bibinfo {author} {\bibfnamefont {S.}~\bibnamefont
  {Gaertner}}, \bibinfo {author} {\bibfnamefont {C.}~\bibnamefont
  {Kurtsiefer}}, \bibinfo {author} {\bibfnamefont {M.}~\bibnamefont
  {Bourennane}},\ and\ \bibinfo {author} {\bibfnamefont {H.}~\bibnamefont
  {Weinfurter}},\ }\bibfield  {title} {\bibinfo {title} {Experimental
  demonstration of four-party quantum secret sharing},\ }\href@noop {}
  {\bibfield  {journal} {\bibinfo  {journal} {Phys. Rev. Lett.}\ }\textbf
  {\bibinfo {volume} {98}},\ \bibinfo {pages} {020503} (\bibinfo {year}
  {2007})}\BibitemShut {NoStop}%
\bibitem [{\citenamefont {Bell}\ \emph {et~al.}(2014)\citenamefont {Bell},
  \citenamefont {Markham}, \citenamefont {Herrera-Mart{\'\i}}, \citenamefont
  {Marin}, \citenamefont {Wadsworth}, \citenamefont {Rarity},\ and\
  \citenamefont {Tame}}]{Bell2014ExperimentGraphQSS}%
  \BibitemOpen
  \bibfield  {author} {\bibinfo {author} {\bibfnamefont {B.}~\bibnamefont
  {Bell}}, \bibinfo {author} {\bibfnamefont {D.}~\bibnamefont {Markham}},
  \bibinfo {author} {\bibfnamefont {D.}~\bibnamefont {Herrera-Mart{\'\i}}},
  \bibinfo {author} {\bibfnamefont {A.}~\bibnamefont {Marin}}, \bibinfo
  {author} {\bibfnamefont {W.}~\bibnamefont {Wadsworth}}, \bibinfo {author}
  {\bibfnamefont {J.}~\bibnamefont {Rarity}},\ and\ \bibinfo {author}
  {\bibfnamefont {M.}~\bibnamefont {Tame}},\ }\bibfield  {title} {\bibinfo
  {title} {Experimental demonstration of graph-state quantum secret sharing},\
  }\href@noop {} {\bibfield  {journal} {\bibinfo  {journal} {Nat. Commun.}\
  }\textbf {\bibinfo {volume} {5}},\ \bibinfo {pages} {5480} (\bibinfo {year}
  {2014})}\BibitemShut {NoStop}%
\bibitem [{\citenamefont {Lu}\ \emph {et~al.}(2016)\citenamefont {Lu},
  \citenamefont {Zhang}, \citenamefont {Chen}, \citenamefont {Li},
  \citenamefont {Liu}, \citenamefont {Li}, \citenamefont {Liu}, \citenamefont
  {Ma}, \citenamefont {Chen},\ and\ \citenamefont {Pan}}]{lu2016secret}%
  \BibitemOpen
  \bibfield  {author} {\bibinfo {author} {\bibfnamefont {H.}~\bibnamefont
  {Lu}}, \bibinfo {author} {\bibfnamefont {Z.}~\bibnamefont {Zhang}}, \bibinfo
  {author} {\bibfnamefont {L.-K.}\ \bibnamefont {Chen}}, \bibinfo {author}
  {\bibfnamefont {Z.-D.}\ \bibnamefont {Li}}, \bibinfo {author} {\bibfnamefont
  {C.}~\bibnamefont {Liu}}, \bibinfo {author} {\bibfnamefont {L.}~\bibnamefont
  {Li}}, \bibinfo {author} {\bibfnamefont {N.-L.}\ \bibnamefont {Liu}},
  \bibinfo {author} {\bibfnamefont {X.}~\bibnamefont {Ma}}, \bibinfo {author}
  {\bibfnamefont {Y.-A.}\ \bibnamefont {Chen}},\ and\ \bibinfo {author}
  {\bibfnamefont {J.-W.}\ \bibnamefont {Pan}},\ }\bibfield  {title} {\bibinfo
  {title} {Secret sharing of a quantum state},\ }\href@noop {} {\bibfield
  {journal} {\bibinfo  {journal} {Phys. Rev. Lett.}\ }\textbf {\bibinfo
  {volume} {117}},\ \bibinfo {pages} {030501} (\bibinfo {year}
  {2016})}\BibitemShut {NoStop}%
\bibitem [{\citenamefont {Zhou}\ \emph {et~al.}(2018)\citenamefont {Zhou},
  \citenamefont {Yu}, \citenamefont {Yan}, \citenamefont {Jia}, \citenamefont
  {Zhang}, \citenamefont {Xie},\ and\ \citenamefont
  {Peng}}]{Zhou2018BoundEntangleQSS}%
  \BibitemOpen
  \bibfield  {author} {\bibinfo {author} {\bibfnamefont {Y.}~\bibnamefont
  {Zhou}}, \bibinfo {author} {\bibfnamefont {J.}~\bibnamefont {Yu}}, \bibinfo
  {author} {\bibfnamefont {Z.}~\bibnamefont {Yan}}, \bibinfo {author}
  {\bibfnamefont {X.}~\bibnamefont {Jia}}, \bibinfo {author} {\bibfnamefont
  {J.}~\bibnamefont {Zhang}}, \bibinfo {author} {\bibfnamefont
  {C.}~\bibnamefont {Xie}},\ and\ \bibinfo {author} {\bibfnamefont
  {K.}~\bibnamefont {Peng}},\ }\bibfield  {title} {\bibinfo {title} {Quantum
  secret sharing among four players using multipartite bound entanglement of an
  optical field},\ }\href@noop {} {\bibfield  {journal} {\bibinfo  {journal}
  {Phys. Rev. Lett.}\ }\textbf {\bibinfo {volume} {121}},\ \bibinfo {pages}
  {150502} (\bibinfo {year} {2018})}\BibitemShut {NoStop}%
\bibitem [{\citenamefont {Williams}\ \emph {et~al.}(2019)\citenamefont
  {Williams}, \citenamefont {Lukens}, \citenamefont {Peters}, \citenamefont
  {Qi},\ and\ \citenamefont {Grice}}]{Williams2019EntangleQSS}%
  \BibitemOpen
  \bibfield  {author} {\bibinfo {author} {\bibfnamefont {B.~P.}\ \bibnamefont
  {Williams}}, \bibinfo {author} {\bibfnamefont {J.~M.}\ \bibnamefont
  {Lukens}}, \bibinfo {author} {\bibfnamefont {N.~A.}\ \bibnamefont {Peters}},
  \bibinfo {author} {\bibfnamefont {B.}~\bibnamefont {Qi}},\ and\ \bibinfo
  {author} {\bibfnamefont {W.~P.}\ \bibnamefont {Grice}},\ }\bibfield  {title}
  {\bibinfo {title} {Quantum secret sharing with polarization-entangled photon
  pairs},\ }\href@noop {} {\bibfield  {journal} {\bibinfo  {journal} {Phys.
  Rev. A}\ }\textbf {\bibinfo {volume} {99}},\ \bibinfo {pages} {062311}
  (\bibinfo {year} {2019})}\BibitemShut {NoStop}%
\bibitem [{\citenamefont {Lee}\ \emph {et~al.}(2020)\citenamefont {Lee},
  \citenamefont {Lee}, \citenamefont {Jeong},\ and\ \citenamefont
  {Park}}]{lee2020quantum}%
  \BibitemOpen
  \bibfield  {author} {\bibinfo {author} {\bibfnamefont {S.~M.}\ \bibnamefont
  {Lee}}, \bibinfo {author} {\bibfnamefont {S.-W.}\ \bibnamefont {Lee}},
  \bibinfo {author} {\bibfnamefont {H.}~\bibnamefont {Jeong}},\ and\ \bibinfo
  {author} {\bibfnamefont {H.~S.}\ \bibnamefont {Park}},\ }\bibfield  {title}
  {\bibinfo {title} {Quantum teleportation of shared quantum secret},\
  }\href@noop {} {\bibfield  {journal} {\bibinfo  {journal} {Phys. Rev. Lett.}\
  }\textbf {\bibinfo {volume} {124}},\ \bibinfo {pages} {060501} (\bibinfo
  {year} {2020})}\BibitemShut {NoStop}%
\bibitem [{\citenamefont {Liu}\ \emph {et~al.}(2021{\natexlab{a}})\citenamefont
  {Liu}, \citenamefont {Liang}, \citenamefont {Sun}, \citenamefont {Li},
  \citenamefont {Meng}, \citenamefont {Yang}, \citenamefont {Li}, \citenamefont
  {Chen}, \citenamefont {Xu}, \citenamefont {Li} \emph
  {et~al.}}]{liu2021photonic}%
  \BibitemOpen
  \bibfield  {author} {\bibinfo {author} {\bibfnamefont {Z.-H.}\ \bibnamefont
  {Liu}}, \bibinfo {author} {\bibfnamefont {X.-B.}\ \bibnamefont {Liang}},
  \bibinfo {author} {\bibfnamefont {K.}~\bibnamefont {Sun}}, \bibinfo {author}
  {\bibfnamefont {Q.}~\bibnamefont {Li}}, \bibinfo {author} {\bibfnamefont
  {Y.}~\bibnamefont {Meng}}, \bibinfo {author} {\bibfnamefont {M.}~\bibnamefont
  {Yang}}, \bibinfo {author} {\bibfnamefont {B.}~\bibnamefont {Li}}, \bibinfo
  {author} {\bibfnamefont {J.-L.}\ \bibnamefont {Chen}}, \bibinfo {author}
  {\bibfnamefont {J.-S.}\ \bibnamefont {Xu}}, \bibinfo {author} {\bibfnamefont
  {C.-F.}\ \bibnamefont {Li}} \emph {et~al.},\ }\bibfield  {title} {\bibinfo
  {title} {Photonic implementation of quantum information masking},\
  }\href@noop {} {\bibfield  {journal} {\bibinfo  {journal} {Phys. Rev. Lett.}\
  }\textbf {\bibinfo {volume} {126}},\ \bibinfo {pages} {170505} (\bibinfo
  {year} {2021}{\natexlab{a}})}\BibitemShut {NoStop}%
\bibitem [{\citenamefont {Shen}\ \emph {et~al.}(2023)\citenamefont {Shen},
  \citenamefont {Cao}, \citenamefont {Wang}, \citenamefont {Fu}, \citenamefont
  {Gu}, \citenamefont {Liu}, \citenamefont {Weng}, \citenamefont {Yin},\ and\
  \citenamefont {Chen}}]{shen2023experimental}%
  \BibitemOpen
  \bibfield  {author} {\bibinfo {author} {\bibfnamefont {A.}~\bibnamefont
  {Shen}}, \bibinfo {author} {\bibfnamefont {X.-Y.}\ \bibnamefont {Cao}},
  \bibinfo {author} {\bibfnamefont {Y.}~\bibnamefont {Wang}}, \bibinfo {author}
  {\bibfnamefont {Y.}~\bibnamefont {Fu}}, \bibinfo {author} {\bibfnamefont
  {J.}~\bibnamefont {Gu}}, \bibinfo {author} {\bibfnamefont {W.-B.}\
  \bibnamefont {Liu}}, \bibinfo {author} {\bibfnamefont {C.-X.}\ \bibnamefont
  {Weng}}, \bibinfo {author} {\bibfnamefont {H.-L.}\ \bibnamefont {Yin}},\ and\
  \bibinfo {author} {\bibfnamefont {Z.-B.}\ \bibnamefont {Chen}},\ }\bibfield
  {title} {\bibinfo {title} {Experimental quantum secret sharing based on phase
  encoding of coherent states},\ }\href@noop {} {\bibfield  {journal} {\bibinfo
   {journal} {Sci. China-Phys. Mech. Astron.}\ }\textbf {\bibinfo {volume}
  {66}},\ \bibinfo {pages} {260311} (\bibinfo {year} {2023})}\BibitemShut
  {NoStop}%
\bibitem [{\citenamefont {Liu}\ \emph {et~al.}(2023)\citenamefont {Liu},
  \citenamefont {Lu}, \citenamefont {Wang}, \citenamefont {Tian}, \citenamefont
  {Wang},\ and\ \citenamefont {Li}}]{liu2023experimental}%
  \BibitemOpen
  \bibfield  {author} {\bibinfo {author} {\bibfnamefont {S.}~\bibnamefont
  {Liu}}, \bibinfo {author} {\bibfnamefont {Z.}~\bibnamefont {Lu}}, \bibinfo
  {author} {\bibfnamefont {P.}~\bibnamefont {Wang}}, \bibinfo {author}
  {\bibfnamefont {Y.}~\bibnamefont {Tian}}, \bibinfo {author} {\bibfnamefont
  {X.}~\bibnamefont {Wang}},\ and\ \bibinfo {author} {\bibfnamefont
  {Y.}~\bibnamefont {Li}},\ }\bibfield  {title} {\bibinfo {title} {Experimental
  demonstration of multiparty quantum secret sharing and conference key
  agreement},\ }\href@noop {} {\bibfield  {journal} {\bibinfo  {journal} {npj
  Quantum Inf.}\ }\textbf {\bibinfo {volume} {9}},\ \bibinfo {pages} {92}
  (\bibinfo {year} {2023})}\BibitemShut {NoStop}%
\bibitem [{\citenamefont {Qin}\ \emph {et~al.}(2024)\citenamefont {Qin},
  \citenamefont {Cheng}, \citenamefont {Ma}, \citenamefont {Zhao},
  \citenamefont {Yan}, \citenamefont {Jia}, \citenamefont {Xie},\ and\
  \citenamefont {Peng}}]{qin2024efficient}%
  \BibitemOpen
  \bibfield  {author} {\bibinfo {author} {\bibfnamefont {Y.}~\bibnamefont
  {Qin}}, \bibinfo {author} {\bibfnamefont {J.}~\bibnamefont {Cheng}}, \bibinfo
  {author} {\bibfnamefont {J.}~\bibnamefont {Ma}}, \bibinfo {author}
  {\bibfnamefont {D.}~\bibnamefont {Zhao}}, \bibinfo {author} {\bibfnamefont
  {Z.}~\bibnamefont {Yan}}, \bibinfo {author} {\bibfnamefont {X.}~\bibnamefont
  {Jia}}, \bibinfo {author} {\bibfnamefont {C.}~\bibnamefont {Xie}},\ and\
  \bibinfo {author} {\bibfnamefont {K.}~\bibnamefont {Peng}},\ }\bibfield
  {title} {\bibinfo {title} {Efficient and secure quantum secret sharing for
  eight users},\ }\href@noop {} {\bibfield  {journal} {\bibinfo  {journal}
  {Phys. Rev. Res.}\ }\textbf {\bibinfo {volume} {6}},\ \bibinfo {pages}
  {033036} (\bibinfo {year} {2024})}\BibitemShut {NoStop}%
\bibitem [{\citenamefont {Gottesman}\ and\ \citenamefont
  {Chuang}(1999)}]{gottesman1999demonstrating}%
  \BibitemOpen
  \bibfield  {author} {\bibinfo {author} {\bibfnamefont {D.}~\bibnamefont
  {Gottesman}}\ and\ \bibinfo {author} {\bibfnamefont {I.~L.}\ \bibnamefont
  {Chuang}},\ }\bibfield  {title} {\bibinfo {title} {Demonstrating the
  viability of universal quantum computation using teleportation and
  single-qubit operations},\ }\href@noop {} {\bibfield  {journal} {\bibinfo
  {journal} {Nature (London)}\ }\textbf {\bibinfo {volume} {402}},\ \bibinfo {pages}
  {390} (\bibinfo {year} {1999})}\BibitemShut {NoStop}%
\bibitem [{\citenamefont {Cao}\ \emph {et~al.}(2024)\citenamefont {Cao},
  \citenamefont {Li}, \citenamefont {Wang}, \citenamefont {Fu}, \citenamefont
  {Yin},\ and\ \citenamefont {Chen}}]{Cao2024Qecommerce}%
  \BibitemOpen
  \bibfield  {author} {\bibinfo {author} {\bibfnamefont {X.-Y.}\ \bibnamefont
  {Cao}}, \bibinfo {author} {\bibfnamefont {B.-H.}\ \bibnamefont {Li}},
  \bibinfo {author} {\bibfnamefont {Y.}~\bibnamefont {Wang}}, \bibinfo {author}
  {\bibfnamefont {Y.}~\bibnamefont {Fu}}, \bibinfo {author} {\bibfnamefont
  {H.-L.}\ \bibnamefont {Yin}},\ and\ \bibinfo {author} {\bibfnamefont {Z.-B.}\
  \bibnamefont {Chen}},\ }\bibfield  {title} {\bibinfo {title} {Experimental
  quantum e-commerce},\ }\href@noop {} {\bibfield  {journal} {\bibinfo
  {journal} {Sci. Adv.}\ }\textbf {\bibinfo {volume} {10}},\ \bibinfo {pages}
  {eadk3258} (\bibinfo {year} {2024})}\BibitemShut {NoStop}%
\bibitem [{\citenamefont {Jing}\ \emph {et~al.}(2024)\citenamefont {Jing},
  \citenamefont {Qian}, \citenamefont {Weng}, \citenamefont {Li}, \citenamefont
  {Chen}, \citenamefont {Wang}, \citenamefont {Tang}, \citenamefont {Gu},
  \citenamefont {Kong}, \citenamefont {Chen} \emph
  {et~al.}}]{jing2024experimental}%
  \BibitemOpen
  \bibfield  {author} {\bibinfo {author} {\bibfnamefont {X.}~\bibnamefont
  {Jing}}, \bibinfo {author} {\bibfnamefont {C.}~\bibnamefont {Qian}}, \bibinfo
  {author} {\bibfnamefont {C.-X.}\ \bibnamefont {Weng}}, \bibinfo {author}
  {\bibfnamefont {B.-H.}\ \bibnamefont {Li}}, \bibinfo {author} {\bibfnamefont
  {Z.}~\bibnamefont {Chen}}, \bibinfo {author} {\bibfnamefont {C.-Q.}\
  \bibnamefont {Wang}}, \bibinfo {author} {\bibfnamefont {J.}~\bibnamefont
  {Tang}}, \bibinfo {author} {\bibfnamefont {X.-W.}\ \bibnamefont {Gu}},
  \bibinfo {author} {\bibfnamefont {Y.-C.}\ \bibnamefont {Kong}}, \bibinfo
  {author} {\bibfnamefont {T.-S.}\ \bibnamefont {Chen}} \emph {et~al.},\
  }\bibfield  {title} {\bibinfo {title} {Experimental quantum {B}yzantine
  agreement on a three-user quantum network with integrated photonics},\
  }\href@noop {} {\bibfield  {journal} {\bibinfo  {journal} {Sci. Adv.}\
  }\textbf {\bibinfo {volume} {10}},\ \bibinfo {pages} {eadp2877} (\bibinfo
  {year} {2024})}\BibitemShut {NoStop}%
\bibitem [{\citenamefont {Yao}\ \emph {et~al.}(2012)\citenamefont {Yao},
  \citenamefont {Wang}, \citenamefont {Xu}, \citenamefont {Lu}, \citenamefont
  {Pan}, \citenamefont {Bao}, \citenamefont {Peng}, \citenamefont {Lu},
  \citenamefont {Chen},\ and\ \citenamefont {Pan}}]{yao2012observation}%
  \BibitemOpen
  \bibfield  {author} {\bibinfo {author} {\bibfnamefont {X.-C.}\ \bibnamefont
  {Yao}}, \bibinfo {author} {\bibfnamefont {T.-X.}\ \bibnamefont {Wang}},
  \bibinfo {author} {\bibfnamefont {P.}~\bibnamefont {Xu}}, \bibinfo {author}
  {\bibfnamefont {H.}~\bibnamefont {Lu}}, \bibinfo {author} {\bibfnamefont
  {G.-S.}\ \bibnamefont {Pan}}, \bibinfo {author} {\bibfnamefont {X.-H.}\
  \bibnamefont {Bao}}, \bibinfo {author} {\bibfnamefont {C.-Z.}\ \bibnamefont
  {Peng}}, \bibinfo {author} {\bibfnamefont {C.-Y.}\ \bibnamefont {Lu}},
  \bibinfo {author} {\bibfnamefont {Y.-A.}\ \bibnamefont {Chen}},\ and\
  \bibinfo {author} {\bibfnamefont {J.-W.}\ \bibnamefont {Pan}},\ }\bibfield
  {title} {\bibinfo {title} {Observation of eight-photon entanglement},\
  }\href@noop {} {\bibfield  {journal} {\bibinfo  {journal} {Nat. Photonics}\
  }\textbf {\bibinfo {volume} {6}},\ \bibinfo {pages} {225} (\bibinfo {year}
  {2012})}\BibitemShut {NoStop}%
\bibitem [{\citenamefont {Wang}\ \emph {et~al.}(2016)\citenamefont {Wang},
  \citenamefont {Chen}, \citenamefont {Li}, \citenamefont {Huang},
  \citenamefont {Liu}, \citenamefont {Chen}, \citenamefont {Luo}, \citenamefont
  {Su}, \citenamefont {Wu}, \citenamefont {Li} \emph
  {et~al.}}]{wang2016experimental}%
  \BibitemOpen
  \bibfield  {author} {\bibinfo {author} {\bibfnamefont {X.-L.}\ \bibnamefont
  {Wang}}, \bibinfo {author} {\bibfnamefont {L.-K.}\ \bibnamefont {Chen}},
  \bibinfo {author} {\bibfnamefont {W.}~\bibnamefont {Li}}, \bibinfo {author}
  {\bibfnamefont {H.-L.}\ \bibnamefont {Huang}}, \bibinfo {author}
  {\bibfnamefont {C.}~\bibnamefont {Liu}}, \bibinfo {author} {\bibfnamefont
  {C.}~\bibnamefont {Chen}}, \bibinfo {author} {\bibfnamefont {Y.-H.}\
  \bibnamefont {Luo}}, \bibinfo {author} {\bibfnamefont {Z.-E.}\ \bibnamefont
  {Su}}, \bibinfo {author} {\bibfnamefont {D.}~\bibnamefont {Wu}}, \bibinfo
  {author} {\bibfnamefont {Z.-D.}\ \bibnamefont {Li}} \emph {et~al.},\
  }\bibfield  {title} {\bibinfo {title} {Experimental ten-photon
  entanglement},\ }\href@noop {} {\bibfield  {journal} {\bibinfo  {journal}
  {Phys. Rev. Lett.}\ }\textbf {\bibinfo {volume} {117}},\ \bibinfo {pages}
  {210502} (\bibinfo {year} {2016})}\BibitemShut {NoStop}%
\bibitem [{\citenamefont {Zhong}\ \emph {et~al.}(2018)\citenamefont {Zhong},
  \citenamefont {Li}, \citenamefont {Li}, \citenamefont {Peng}, \citenamefont
  {Su}, \citenamefont {Hu}, \citenamefont {He}, \citenamefont {Ding},
  \citenamefont {Zhang}, \citenamefont {Li} \emph {et~al.}}]{zhong201812}%
  \BibitemOpen
  \bibfield  {author} {\bibinfo {author} {\bibfnamefont {H.-S.}\ \bibnamefont
  {Zhong}}, \bibinfo {author} {\bibfnamefont {Y.}~\bibnamefont {Li}}, \bibinfo
  {author} {\bibfnamefont {W.}~\bibnamefont {Li}}, \bibinfo {author}
  {\bibfnamefont {L.-C.}\ \bibnamefont {Peng}}, \bibinfo {author}
  {\bibfnamefont {Z.-E.}\ \bibnamefont {Su}}, \bibinfo {author} {\bibfnamefont
  {Y.}~\bibnamefont {Hu}}, \bibinfo {author} {\bibfnamefont {Y.-M.}\
  \bibnamefont {He}}, \bibinfo {author} {\bibfnamefont {X.}~\bibnamefont
  {Ding}}, \bibinfo {author} {\bibfnamefont {W.}~\bibnamefont {Zhang}},
  \bibinfo {author} {\bibfnamefont {H.}~\bibnamefont {Li}} \emph {et~al.},\
  }\bibfield  {title} {\bibinfo {title} {12-photon entanglement and scalable
  scattershot boson sampling with optimal entangled-photon pairs from
  parametric down-conversion},\ }\href@noop {} {\bibfield  {journal} {\bibinfo
  {journal} {Phys. Rev. Lett.}\ }\textbf {\bibinfo {volume} {121}},\ \bibinfo
  {pages} {250505} (\bibinfo {year} {2018})}\BibitemShut {NoStop}%
\bibitem [{\citenamefont {Karlsson}\ \emph {et~al.}(1999)\citenamefont
  {Karlsson}, \citenamefont {Koashi},\ and\ \citenamefont
  {Imoto}}]{karlsson1999quantum}%
  \BibitemOpen
  \bibfield  {author} {\bibinfo {author} {\bibfnamefont {A.}~\bibnamefont
  {Karlsson}}, \bibinfo {author} {\bibfnamefont {M.}~\bibnamefont {Koashi}},\
  and\ \bibinfo {author} {\bibfnamefont {N.}~\bibnamefont {Imoto}},\ }\bibfield
   {title} {\bibinfo {title} {Quantum entanglement for secret sharing and
  secret splitting},\ }\href@noop {} {\bibfield  {journal} {\bibinfo  {journal}
  {Phys. Rev. A}\ }\textbf {\bibinfo {volume} {59}},\ \bibinfo {pages} {162}
  (\bibinfo {year} {1999})}\BibitemShut {NoStop}%
\bibitem [{\citenamefont {Qin}\ \emph {et~al.}(2007)\citenamefont {Qin},
  \citenamefont {Gao}, \citenamefont {Wen},\ and\ \citenamefont
  {Zhu}}]{qin2007cryptanalysis}%
  \BibitemOpen
  \bibfield  {author} {\bibinfo {author} {\bibfnamefont {S.-J.}\ \bibnamefont
  {Qin}}, \bibinfo {author} {\bibfnamefont {F.}~\bibnamefont {Gao}}, \bibinfo
  {author} {\bibfnamefont {Q.-Y.}\ \bibnamefont {Wen}},\ and\ \bibinfo {author}
  {\bibfnamefont {F.-C.}\ \bibnamefont {Zhu}},\ }\bibfield  {title} {\bibinfo
  {title} {Cryptanalysis of the {H}illery-{B}u{\v{z}}ek-{B}erthiaume quantum
  secret-sharing protocol},\ }\href@noop {} {\bibfield  {journal} {\bibinfo
  {journal} {Phys. Rev. A}\ }\textbf {\bibinfo {volume} {76}},\ \bibinfo
  {pages} {062324} (\bibinfo {year} {2007})}\BibitemShut {NoStop}%
\bibitem [{\citenamefont {Xiao}\ \emph {et~al.}(2024)\citenamefont {Xiao},
  \citenamefont {Jia}, \citenamefont {Song}, \citenamefont {Bao}, \citenamefont
  {Fu}, \citenamefont {Yin},\ and\ \citenamefont {Chen}}]{xiao2024source}%
  \BibitemOpen
  \bibfield  {author} {\bibinfo {author} {\bibfnamefont {Y.-R.}\ \bibnamefont
  {Xiao}}, \bibinfo {author} {\bibfnamefont {Z.-Y.}\ \bibnamefont {Jia}},
  \bibinfo {author} {\bibfnamefont {Y.-C.}\ \bibnamefont {Song}}, \bibinfo
  {author} {\bibfnamefont {Y.}~\bibnamefont {Bao}}, \bibinfo {author}
  {\bibfnamefont {Y.}~\bibnamefont {Fu}}, \bibinfo {author} {\bibfnamefont
  {H.-L.}\ \bibnamefont {Yin}},\ and\ \bibinfo {author} {\bibfnamefont {Z.-B.}\
  \bibnamefont {Chen}},\ }\bibfield  {title} {\bibinfo {title}
  {Source-independent quantum secret sharing with entangled photon pair
  networks},\ }\href@noop {} {\bibfield  {journal} {\bibinfo  {journal} {Opt.
  Lett.}\ }\textbf {\bibinfo {volume} {49}},\ \bibinfo {pages} {4210} (\bibinfo
  {year} {2024})}\BibitemShut {NoStop}%
\bibitem [{\citenamefont {Lu}\ \emph {et~al.}(2021)\citenamefont {Lu},
  \citenamefont {Cao}, \citenamefont {Weng}, \citenamefont {Gu}, \citenamefont
  {Xie}, \citenamefont {Zhou}, \citenamefont {Yin},\ and\ \citenamefont
  {Chen}}]{Lu2021EfficientQDS}%
  \BibitemOpen
  \bibfield  {author} {\bibinfo {author} {\bibfnamefont {Y.-S.}\ \bibnamefont
  {Lu}}, \bibinfo {author} {\bibfnamefont {X.-Y.}\ \bibnamefont {Cao}},
  \bibinfo {author} {\bibfnamefont {C.-X.}\ \bibnamefont {Weng}}, \bibinfo
  {author} {\bibfnamefont {J.}~\bibnamefont {Gu}}, \bibinfo {author}
  {\bibfnamefont {Y.-M.}\ \bibnamefont {Xie}}, \bibinfo {author} {\bibfnamefont
  {M.-G.}\ \bibnamefont {Zhou}}, \bibinfo {author} {\bibfnamefont {H.-L.}\
  \bibnamefont {Yin}},\ and\ \bibinfo {author} {\bibfnamefont {Z.-B.}\
  \bibnamefont {Chen}},\ }\bibfield  {title} {\bibinfo {title} {Efficient
  quantum digital signatures without symmetrization step},\ }\href@noop {}
  {\bibfield  {journal} {\bibinfo  {journal} {Opt. Express}\ }\textbf {\bibinfo
  {volume} {29}},\ \bibinfo {pages} {10162} (\bibinfo {year}
  {2021})}\BibitemShut {NoStop}%
\bibitem [{\citenamefont {Walk}\ and\ \citenamefont
  {Eisert}(2021)}]{Walk2021Sharing}%
  \BibitemOpen
  \bibfield  {author} {\bibinfo {author} {\bibfnamefont {N.}~\bibnamefont
  {Walk}}\ and\ \bibinfo {author} {\bibfnamefont {J.}~\bibnamefont {Eisert}},\
  }\bibfield  {title} {\bibinfo {title} {Sharing classical secrets with
  continuous-variable entanglement: Composable security and network coding
  advantage},\ }\href@noop {} {\bibfield  {journal} {\bibinfo  {journal} {PRX
  Quantum}\ }\textbf {\bibinfo {volume} {2}},\ \bibinfo {pages} {040339}
  (\bibinfo {year} {2021})}\BibitemShut {NoStop}%
\bibitem [{\citenamefont {Kim}\ \emph {et~al.}(2006)\citenamefont {Kim},
  \citenamefont {Fiorentino},\ and\ \citenamefont {Wong}}]{kim2006phase}%
  \BibitemOpen
  \bibfield  {author} {\bibinfo {author} {\bibfnamefont {T.}~\bibnamefont
  {Kim}}, \bibinfo {author} {\bibfnamefont {M.}~\bibnamefont {Fiorentino}},\
  and\ \bibinfo {author} {\bibfnamefont {F.~N.~C.}\ \bibnamefont {Wong}},\
  }\bibfield  {title} {\bibinfo {title} {Phase-stable source of
  polarization-entangled photons using a polarization sagnac interferometer},\
  }\href@noop {} {\bibfield  {journal} {\bibinfo  {journal} {Phys. Rev. A}\
  }\textbf {\bibinfo {volume} {73}},\ \bibinfo {pages} {012316} (\bibinfo
  {year} {2006})}\BibitemShut {NoStop}%
\bibitem [{\citenamefont {Fedrizzi}\ \emph {et~al.}(2007)\citenamefont
  {Fedrizzi}, \citenamefont {Herbst}, \citenamefont {Poppe}, \citenamefont
  {Jennewein},\ and\ \citenamefont {Zeilinger}}]{fedrizzi2007wavelength}%
  \BibitemOpen
  \bibfield  {author} {\bibinfo {author} {\bibfnamefont {A.}~\bibnamefont
  {Fedrizzi}}, \bibinfo {author} {\bibfnamefont {T.}~\bibnamefont {Herbst}},
  \bibinfo {author} {\bibfnamefont {A.}~\bibnamefont {Poppe}}, \bibinfo
  {author} {\bibfnamefont {T.}~\bibnamefont {Jennewein}},\ and\ \bibinfo
  {author} {\bibfnamefont {A.}~\bibnamefont {Zeilinger}},\ }\bibfield  {title}
  {\bibinfo {title} {A wavelength-tunable fiber-coupled source of narrowband
  entangled photons},\ }\href@noop {} {\bibfield  {journal} {\bibinfo
  {journal} {Opt. Express}\ }\textbf {\bibinfo {volume} {15}},\ \bibinfo
  {pages} {15377} (\bibinfo {year} {2007})}\BibitemShut {NoStop}%
\bibitem [{\citenamefont {Lim}\ \emph {et~al.}(2008)\citenamefont {Lim},
  \citenamefont {Yoshizawa}, \citenamefont {Tsuchida},\ and\ \citenamefont
  {Kikuchi}}]{lim2008broadband}%
  \BibitemOpen
  \bibfield  {author} {\bibinfo {author} {\bibfnamefont {H.~C.}\ \bibnamefont
  {Lim}}, \bibinfo {author} {\bibfnamefont {A.}~\bibnamefont {Yoshizawa}},
  \bibinfo {author} {\bibfnamefont {H.}~\bibnamefont {Tsuchida}},\ and\
  \bibinfo {author} {\bibfnamefont {K.}~\bibnamefont {Kikuchi}},\ }\bibfield
  {title} {\bibinfo {title} {Broadband source of telecom-band
  polarization-entangled photon-pairs for wavelength-multiplexed entanglement
  distribution},\ }\href@noop {} {\bibfield  {journal} {\bibinfo  {journal}
  {Opt. Express}\ }\textbf {\bibinfo {volume} {16}},\ \bibinfo {pages} {16052}
  (\bibinfo {year} {2008})}\BibitemShut {NoStop}%
\bibitem [{\citenamefont {Fung}\ \emph {et~al.}(2011)\citenamefont {Fung},
  \citenamefont {Chau},\ and\ \citenamefont {Lo}}]{fung2011universal}%
  \BibitemOpen
  \bibfield  {author} {\bibinfo {author} {\bibfnamefont {C.-H.~F.}\
  \bibnamefont {Fung}}, \bibinfo {author} {\bibfnamefont {H.~F.}~\bibnamefont
  {Chau}},\ and\ \bibinfo {author} {\bibfnamefont {H.-K.}\ \bibnamefont {Lo}},\
  }\bibfield  {title} {\bibinfo {title} {Universal squash model for optical
  communications using linear optics and threshold detectors},\ }\href@noop {}
  {\bibfield  {journal} {\bibinfo  {journal} {Phys. Rev. A}\ }\textbf {\bibinfo
  {volume} {84}},\ \bibinfo {pages} {020303(R)} (\bibinfo {year}
  {2011})}\BibitemShut {NoStop}%
\bibitem [{\citenamefont {James}\ \emph {et~al.}(2001)\citenamefont {James},
  \citenamefont {Kwiat}, \citenamefont {Munro},\ and\ \citenamefont
  {White}}]{james2001measurement}%
  \BibitemOpen
  \bibfield  {author} {\bibinfo {author} {\bibfnamefont {D.~F.~V.}\ \bibnamefont
  {James}}, \bibinfo {author} {\bibfnamefont {P.~G.}\ \bibnamefont {Kwiat}},
  \bibinfo {author} {\bibfnamefont {W.~J.}\ \bibnamefont {Munro}},\ and\
  \bibinfo {author} {\bibfnamefont {A.~G.}\ \bibnamefont {White}},\ }\bibfield
  {title} {\bibinfo {title} {Measurement of qubits},\ }\href@noop {} {\bibfield
   {journal} {\bibinfo  {journal} {Phys. Rev. A}\ }\textbf {\bibinfo {volume}
  {64}},\ \bibinfo {pages} {052312} (\bibinfo {year} {2001})}\BibitemShut
  {NoStop}%
\bibitem [{\citenamefont {Yin}\ \emph {et~al.}(2020)\citenamefont {Yin},
  \citenamefont {Zhou}, \citenamefont {Gu}, \citenamefont {Xie}, \citenamefont
  {Lu},\ and\ \citenamefont {Chen}}]{yin2020tight}%
  \BibitemOpen
  \bibfield  {author} {\bibinfo {author} {\bibfnamefont {H.-L.}\ \bibnamefont
  {Yin}}, \bibinfo {author} {\bibfnamefont {M.-G.}\ \bibnamefont {Zhou}},
  \bibinfo {author} {\bibfnamefont {J.}~\bibnamefont {Gu}}, \bibinfo {author}
  {\bibfnamefont {Y.-M.}\ \bibnamefont {Xie}}, \bibinfo {author} {\bibfnamefont
  {Y.-S.}\ \bibnamefont {Lu}},\ and\ \bibinfo {author} {\bibfnamefont {Z.-B.}\
  \bibnamefont {Chen}},\ }\bibfield  {title} {\bibinfo {title} {Tight security
  bounds for decoy-state quantum key distribution},\ }\href@noop {} {\bibfield
  {journal} {\bibinfo  {journal} {Sci. Rep.}\ }\textbf {\bibinfo {volume}
  {10}},\ \bibinfo {pages} {14312} (\bibinfo {year} {2020})}\BibitemShut
  {NoStop}%
\bibitem [{\citenamefont {Liu}\ \emph {et~al.}(2021{\natexlab{b}})\citenamefont
  {Liu}, \citenamefont {Li}, \citenamefont {Ragy}, \citenamefont {Zhao},
  \citenamefont {Bai}, \citenamefont {Liu}, \citenamefont {Brown},
  \citenamefont {Zhang}, \citenamefont {Colbeck}, \citenamefont {Fan} \emph
  {et~al.}}]{liu2021device}%
  \BibitemOpen
  \bibfield  {author} {\bibinfo {author} {\bibfnamefont {W.-Z.}\ \bibnamefont
  {Liu}}, \bibinfo {author} {\bibfnamefont {M.-H.}\ \bibnamefont {Li}},
  \bibinfo {author} {\bibfnamefont {S.}~\bibnamefont {Ragy}}, \bibinfo {author}
  {\bibfnamefont {S.-R.}\ \bibnamefont {Zhao}}, \bibinfo {author}
  {\bibfnamefont {B.}~\bibnamefont {Bai}}, \bibinfo {author} {\bibfnamefont
  {Y.}~\bibnamefont {Liu}}, \bibinfo {author} {\bibfnamefont {P.~J.}\
  \bibnamefont {Brown}}, \bibinfo {author} {\bibfnamefont {J.}~\bibnamefont
  {Zhang}}, \bibinfo {author} {\bibfnamefont {R.}~\bibnamefont {Colbeck}},
  \bibinfo {author} {\bibfnamefont {J.}~\bibnamefont {Fan}} \emph {et~al.},\
  }\bibfield  {title} {\bibinfo {title} {Device-independent randomness
  expansion against quantum side information},\ }\href@noop {} {\bibfield
  {journal} {\bibinfo  {journal} {Nat. Phys.}\ }\textbf {\bibinfo {volume}
  {17}},\ \bibinfo {pages} {448} (\bibinfo {year}
  {2021}{\natexlab{b}})}\BibitemShut {NoStop}%
\bibitem [{\citenamefont {Wengerowsky}\ \emph {et~al.}(2018)\citenamefont
  {Wengerowsky}, \citenamefont {Joshi}, \citenamefont {Steinlechner},
  \citenamefont {H{\"u}bel},\ and\ \citenamefont
  {Ursin}}]{Wengerowsky2018EntangleNetwork}%
  \BibitemOpen
  \bibfield  {author} {\bibinfo {author} {\bibfnamefont {S.}~\bibnamefont
  {Wengerowsky}}, \bibinfo {author} {\bibfnamefont {S.~K.}\ \bibnamefont
  {Joshi}}, \bibinfo {author} {\bibfnamefont {F.}~\bibnamefont {Steinlechner}},
  \bibinfo {author} {\bibfnamefont {H.}~\bibnamefont {H{\"u}bel}},\ and\
  \bibinfo {author} {\bibfnamefont {R.}~\bibnamefont {Ursin}},\ }\bibfield
  {title} {\bibinfo {title} {An entanglement-based wavelength-multiplexed
  quantum communication network},\ }\href@noop {} {\bibfield  {journal}
  {\bibinfo  {journal} {Nature (London)}\ }\textbf {\bibinfo {volume} {564}},\ \bibinfo
  {pages} {225} (\bibinfo {year} {2018})}\BibitemShut {NoStop}%
\bibitem [{\citenamefont {Joshi}\ \emph {et~al.}(2020)\citenamefont {Joshi},
  \citenamefont {Aktas}, \citenamefont {Wengerowsky}, \citenamefont
  {Lon{\v{c}}ari{\'c}}, \citenamefont {Neumann}, \citenamefont {Liu},
  \citenamefont {Scheidl}, \citenamefont {Lorenzo}, \citenamefont {Samec},
  \citenamefont {Kling} \emph {et~al.}}]{Joshi2020EntangleNetwork}%
  \BibitemOpen
  \bibfield  {author} {\bibinfo {author} {\bibfnamefont {S.~K.}\ \bibnamefont
  {Joshi}}, \bibinfo {author} {\bibfnamefont {D.}~\bibnamefont {Aktas}},
  \bibinfo {author} {\bibfnamefont {S.}~\bibnamefont {Wengerowsky}}, \bibinfo
  {author} {\bibfnamefont {M.}~\bibnamefont {Lon{\v{c}}ari{\'c}}}, \bibinfo
  {author} {\bibfnamefont {S.~P.}\ \bibnamefont {Neumann}}, \bibinfo {author}
  {\bibfnamefont {B.}~\bibnamefont {Liu}}, \bibinfo {author} {\bibfnamefont
  {T.}~\bibnamefont {Scheidl}}, \bibinfo {author} {\bibfnamefont {G.~C.}\
  \bibnamefont {Lorenzo}}, \bibinfo {author} {\bibfnamefont
  {{\v{Z}}.}~\bibnamefont {Samec}}, \bibinfo {author} {\bibfnamefont
  {L.}~\bibnamefont {Kling}} \emph {et~al.},\ }\bibfield  {title} {\bibinfo
  {title} {A trusted node--free eight-user metropolitan quantum communication
  network},\ }\href@noop {} {\bibfield  {journal} {\bibinfo  {journal} {Sci.
  Adv.}\ }\textbf {\bibinfo {volume} {6}},\ \bibinfo {pages} {eaba0959}
  (\bibinfo {year} {2020})}\BibitemShut {NoStop}%
\bibitem [{\citenamefont {Liu}\ \emph {et~al.}(2022)\citenamefont {Liu},
  \citenamefont {Liu}, \citenamefont {Xue}, \citenamefont {Wang}, \citenamefont
  {Li}, \citenamefont {Feng}, \citenamefont {Liu}, \citenamefont {Cui},
  \citenamefont {Wang}, \citenamefont {You} \emph {et~al.}}]{liu202240}%
  \BibitemOpen
  \bibfield  {author} {\bibinfo {author} {\bibfnamefont {X.}~\bibnamefont
  {Liu}}, \bibinfo {author} {\bibfnamefont {J.}~\bibnamefont {Liu}}, \bibinfo
  {author} {\bibfnamefont {R.}~\bibnamefont {Xue}}, \bibinfo {author}
  {\bibfnamefont {H.}~\bibnamefont {Wang}}, \bibinfo {author} {\bibfnamefont
  {H.}~\bibnamefont {Li}}, \bibinfo {author} {\bibfnamefont {X.}~\bibnamefont
  {Feng}}, \bibinfo {author} {\bibfnamefont {F.}~\bibnamefont {Liu}}, \bibinfo
  {author} {\bibfnamefont {K.}~\bibnamefont {Cui}}, \bibinfo {author}
  {\bibfnamefont {Z.}~\bibnamefont {Wang}}, \bibinfo {author} {\bibfnamefont
  {L.}~\bibnamefont {You}} \emph {et~al.},\ }\bibfield  {title} {\bibinfo
  {title} {40-user fully connected entanglement-based quantum key distribution
  network without trusted node},\ }\href@noop {} {\bibfield  {journal}
  {\bibinfo  {journal} {PhotoniX}\ }\textbf {\bibinfo {volume} {3}},\ \bibinfo
  {pages} {2} (\bibinfo {year} {2022})}\BibitemShut {NoStop}%
\bibitem [{\citenamefont {Huang}\ \emph {et~al.}(2025)\citenamefont {Huang},
  \citenamefont {Qi}, \citenamefont {Yang}, \citenamefont {Zhang},
  \citenamefont {Li}, \citenamefont {Zheng},\ and\ \citenamefont
  {Chen}}]{huang2025sixteen}%
  \BibitemOpen
  \bibfield  {author} {\bibinfo {author} {\bibfnamefont {Y.}~\bibnamefont
  {Huang}}, \bibinfo {author} {\bibfnamefont {Z.}~\bibnamefont {Qi}}, \bibinfo
  {author} {\bibfnamefont {Y.}~\bibnamefont {Yang}}, \bibinfo {author}
  {\bibfnamefont {Y.}~\bibnamefont {Zhang}}, \bibinfo {author} {\bibfnamefont
  {Y.}~\bibnamefont {Li}}, \bibinfo {author} {\bibfnamefont {Y.}~\bibnamefont
  {Zheng}},\ and\ \bibinfo {author} {\bibfnamefont {X.}~\bibnamefont {Chen}},\
  }\bibfield  {title} {\bibinfo {title} {A sixteen-user time-bin entangled
  quantum communication network with fully connected topology},\ }\href@noop {}
  {\bibfield  {journal} {\bibinfo  {journal} {Laser Photonics Rev.}\ }\textbf
  {\bibinfo {volume} {19}},\ \bibinfo {pages} {2301026} (\bibinfo {year}
  {2025})}\BibitemShut {NoStop}%
\bibitem [{\citenamefont {Terhal}(2004)}]{Terhal2004EntangleMonogamous}%
  \BibitemOpen
  \bibfield  {author} {\bibinfo {author} {\bibfnamefont {B.~M.}\ \bibnamefont
  {Terhal}},\ }\bibfield  {title} {\bibinfo {title} {Is entanglement
  monogamous?},\ }\href@noop {} {\bibfield  {journal} {\bibinfo  {journal} {IBM
  J. Res. Dev.}\ }\textbf {\bibinfo {volume} {48}},\ \bibinfo {pages} {71}
  (\bibinfo {year} {2004})}\BibitemShut {NoStop}%
\bibitem [{\citenamefont {Kogias}\ \emph {et~al.}(2017)\citenamefont {Kogias},
  \citenamefont {Xiang}, \citenamefont {He},\ and\ \citenamefont
  {Adesso}}]{Kogias2017CVQSS}%
  \BibitemOpen
  \bibfield  {author} {\bibinfo {author} {\bibfnamefont {I.}~\bibnamefont
  {Kogias}}, \bibinfo {author} {\bibfnamefont {Y.}~\bibnamefont {Xiang}},
  \bibinfo {author} {\bibfnamefont {Q.}~\bibnamefont {He}},\ and\ \bibinfo
  {author} {\bibfnamefont {G.}~\bibnamefont {Adesso}},\ }\bibfield  {title}
  {\bibinfo {title} {Unconditional security of entanglement-based
  continuous-variable quantum secret sharing},\ }\href@noop {} {\bibfield
  {journal} {\bibinfo  {journal} {Phys. Rev. A}\ }\textbf {\bibinfo {volume}
  {95}},\ \bibinfo {pages} {012315} (\bibinfo {year} {2017})}\BibitemShut
  {NoStop}%
\bibitem [{\citenamefont {Tomamichel}\ \emph {et~al.}(2012)\citenamefont
  {Tomamichel}, \citenamefont {Lim}, \citenamefont {Gisin},\ and\ \citenamefont
  {Renner}}]{tomamichel2012tight}%
  \BibitemOpen
  \bibfield  {author} {\bibinfo {author} {\bibfnamefont {M.}~\bibnamefont
  {Tomamichel}}, \bibinfo {author} {\bibfnamefont {C.~C.~W.}\ \bibnamefont
  {Lim}}, \bibinfo {author} {\bibfnamefont {N.}~\bibnamefont {Gisin}},\ and\
  \bibinfo {author} {\bibfnamefont {R.}~\bibnamefont {Renner}},\ }\bibfield
  {title} {\bibinfo {title} {Tight finite-key analysis for quantum
  cryptography},\ }\href@noop {} {\bibfield  {journal} {\bibinfo  {journal}
  {Nat. Commun.}\ }\textbf {\bibinfo {volume} {3}},\ \bibinfo {pages} {634}
  (\bibinfo {year} {2012})}\BibitemShut {NoStop}%
\bibitem [{\citenamefont {Renner}(2008)}]{renner2008security}%
  \BibitemOpen
  \bibfield  {author} {\bibinfo {author} {\bibfnamefont {R.}~\bibnamefont
  {Renner}},\ }\bibfield  {title} {\bibinfo {title} {Security of quantum key
  distribution},\ }\href@noop {} {\bibfield  {journal} {\bibinfo  {journal}
  {Int. J. Quantum Inf.}\ }\textbf {\bibinfo {volume} {6}},\ \bibinfo {pages}
  {1} (\bibinfo {year} {2008})}\BibitemShut {NoStop}%
\bibitem [{\citenamefont {Tomamichel}\ \emph {et~al.}(2011)\citenamefont
  {Tomamichel}, \citenamefont {Schaffner}, \citenamefont {Smith},\ and\
  \citenamefont {Renner}}]{tomamichel2011leftover}%
  \BibitemOpen
  \bibfield  {author} {\bibinfo {author} {\bibfnamefont {M.}~\bibnamefont
  {Tomamichel}}, \bibinfo {author} {\bibfnamefont {C.}~\bibnamefont
  {Schaffner}}, \bibinfo {author} {\bibfnamefont {A.}~\bibnamefont {Smith}},\
  and\ \bibinfo {author} {\bibfnamefont {R.}~\bibnamefont {Renner}},\
  }\bibfield  {title} {\bibinfo {title} {Leftover hashing against quantum side
  information},\ }\href@noop {} {\bibfield  {journal} {\bibinfo  {journal}
  {IEEE Trans. Inf. Theory}\ }\textbf {\bibinfo {volume} {57}},\ \bibinfo
  {pages} {5524} (\bibinfo {year} {2011})}\BibitemShut {NoStop}%
\bibitem [{\citenamefont {Tomamichel}\ and\ \citenamefont
  {Renner}(2011)}]{tomamichel2011uncertainty}%
  \BibitemOpen
  \bibfield  {author} {\bibinfo {author} {\bibfnamefont {M.}~\bibnamefont
  {Tomamichel}}\ and\ \bibinfo {author} {\bibfnamefont {R.}~\bibnamefont
  {Renner}},\ }\bibfield  {title} {\bibinfo {title} {Uncertainty relation for
  smooth entropies},\ }\href@noop {} {\bibfield  {journal} {\bibinfo  {journal}
  {Phys. Rev. Lett.}\ }\textbf {\bibinfo {volume} {106}},\ \bibinfo {pages}
  {110506} (\bibinfo {year} {2011})}\BibitemShut {NoStop}%
\end{thebibliography}


\end{document}